\documentclass[sn-mathphys,Numbered]{sn-jnl}

\usepackage{graphicx}%
\usepackage{multirow}%
\usepackage{amsmath,amssymb,amsfonts}%
\usepackage{amsthm}%
\usepackage{mathrsfs}%
\usepackage[title]{appendix}%
\usepackage{xcolor}%
\usepackage{textcomp}%
\usepackage{manyfoot}%
\usepackage{booktabs}%
\usepackage{algorithm}%
\usepackage{algorithmicx}%
\usepackage{algpseudocode}%
\usepackage{listings}%
\usepackage{setspace}
\usepackage{lineno}%
\usepackage{tikz}%
\usepackage{float}%
\usetikzlibrary{decorations.pathreplacing}%
\usepackage{placeins}

\usepackage{changes}

\nolinenumbers

\theoremstyle{thmstyleone}%
\theoremstyle{thmstyletwo}%

\theoremstyle{thmstylethree}%

\begin{document}

\title[Article Title]{The Half-Transform Ansatz: Quarkonium Dynamics in Quantum Phase Space Representation}


\author*[]{\fnm{Gabriel} \sur{Nowaskie}}\email{gabriel.nowaskie341@topper.wku.edu}

\affil*[]{\orgdiv{Department of Physics and Astronomy}, \orgname{The Gatton Academy, Western Kentucky University}, \orgaddress{\street{1906 College Heights Blvd.}, \city{Bowling Green}, \postcode{42101}, \state{KY}, \country{United States}}}


\abstract{Since the groundwork published by Torres-Vega and Frederick, the Quantum Phase Space Representation (QPSR) has been explored as a method for solving a multitude of physical systems and describing phenomena. Most recently, Valentino A. Simpao has developed a method, the Heaviside Operational Ansatz, to solve the Time Dependent Schrodinger Equation (TDSE) in the QPSR, but there are still no general, direct methods to solve the Time Independent Schrodinger Equation in the QPSR. There is also no current formulation of quarkonium in phase space. In this paper, we describe the strong interactions of non-relativistic heavy quarks using the Cornell potential, and present a method, the Half-Transform Ansatz, to cast the Schrodinger Equation into a hyper-geometric form which can be solved for the phase space wave function and its energy eigenvalues using the Nikiforov-Uvarov method. This solution can be generalized for any two particle system with a scleronomic potential made up of polynomial and reciprocal terms. These results are compared to experimental results and other theoretical models. We also analyze the behavior of these wave functions, which suggest a correlation between radial momentum and the upper limit of existence in charm-anticharm mesons.}




\maketitle

\section{Introduction}
\label{sec:introduction}

After the discovery of the $J/\Psi$ meson in 1974 [1], potential models became widely used to describe heavy quark systems. The $J/\Psi$ meson has a mass approximately 3.5 times that of a proton and is the lowest bound state of a charm and anti-charm quark, giving it non-relativistic properties considering its speed to mass ratio. This is observed in charm and bottom quarks, and allows us to use the Schrodinger equation to describe heavy quark interactions.
The non-relativistic nature of heavy quarks is used to imply a static potential, the same as non-relativistic hydrogen atom models. One of the most prominent models, the Cornell potential, characterizes a system by a linear combination of linear and Coulomb potentials [2]. It is written as
\begin{equation}
    \begin{aligned}
        V(r)=\frac{a}{r}+br,
    \end{aligned}
\end{equation}

where the reciprocal term describes the gluon exchange interaction between the quark and anti-quark at short distances. This is asymptotic freedom, and it explains that the strong interaction coupling constant is a function of momentum transfer. At short distances, the momentum transfer increases with quark-antiquark collisions and the coupling constant becomes negligible. This makes the quarks act approximately free. Over longer distances, the momentum transfer decreases and the coupling constant increases. At extreme distances, this results in the quark and anti-quark being confined as described by the linear term. \\


With the discovery of the QPSR by Torres-Vega and Frederick [3], a framework with the capabilities to describe quantum systems in phase space, we can attempt to model quarks in phase space. And the HOA was created and published in 2004 by Simpao to generate exact analytical wave functions in the QPSR.The central developments in the HOA from Simpao can be seen in \ref{section:AppendixA}, which includes additional supporting references. Using the outline created by Simpao [4, 5], we could find the time evolution of such systems. The HOA yields implicit solutions to the Time Independent Schrodinger Equation (TISE) by Fourier transform from the time to the energy domain; however, there are no explicit solution schemes to solving the TISE in the QPSR. In the present work, we march forward on a new line of reasoning, independent of Simpao in [4, 5], to formulate an explicit solution to the TISE without invoking transforms from the time to the energy domain as previously posited by Simpao. There is also no current found full Cornell potential 3D phase space wave function describing quarkonium. \\

This work aims to find a method to solve the TISE in the QPSR for non-relativistic quarkonium and analyze these solutions, and then to generalize these results to give a general method of solving any two particle system with a potential made up of polynomial and reciprocal terms. This work is organized as follows. In section 2, we recap the Heaviside Operational Ansatz (HOA) [4, 5] and construct the Hamiltonian Operator in QPSR. In section 3, we solve the Schrodinger Equation in 1D for a confined, ground state, heavy, non-relativistic, quark-antiquark meson. In section 4, we present a summary of the NU method. In section 5, we present a scheme to find the wave function and complex energy eigenvalues for a quark-antiquark system in 3D. In section 6, we compare our complex energy eigenvalue results with experimental data on heavy quark mass spectra. In section 7, we analyze these mass spectra results. In section 8, we analyze the wave functions to create an ansatz which allows us to find a purely real TISE with real energy eigenvalues. We then compare these new energy eigenvalues with the experimental data. In section 9, we show example charm-anticharm wave function evaluations and analyze the results. In section 10, we conclude our results. In section 1A, we present an appendix for more insight into the HOA and the construction of the Hamiltonian. In section 2A, we present an appendix for more insight into the time dynamics of the 1D wave function.

\bigskip

\section{Recap of the HOA and Hamiltonian Construction}
\label{section:section2}

To begin, we must revisit the Heaviside Operational Ansatz. Only the brief methods relevant to this work are presented. Supplemental developments, historical context, and supporting references are provided in Appendix \ref{section:AppendixA}.  The variable relations: $x,p,t$ are the configuration space position, momentum, and time respectively. When transforming to the QPSR we recall that (where \ $\hat{}$ \ denotes operators and $\alpha,\gamma$ are otherwise free parameters as in [3]): 
\begin{equation*}
    \begin{aligned}
        H\left(x,\ p,\ t\right)\rightarrow\ \hat{H}\left(\hat{x},\ \hat{p},\ t\right)=\hat{H}\left(i\hbar\partial_p+\alpha x,\ -i\hbar\partial_x+\gamma p,\ t\right),\ \ni\alpha+\gamma=1
    \end{aligned}
\end{equation*}
\begin{equation*}
    \begin{aligned}
       x\rightarrow\ \hat{x}\equiv\ i\hbar\partial_p+\alpha\ x,\ p\rightarrow\hat{p}\equiv -i\hbar\partial_x+\gamma\ p,t\rightarrow\ t=t 
    \end{aligned}
\end{equation*}
\begin{equation*}
    \begin{aligned}
       \left(x_1,\ldots,x_n\right)\rightarrow\left(\widehat{x_1},\ldots,\widehat{x_n}\right)= \left(i\hbar\partial_{p_1}+\alpha_1x_1,\ldots,i\hbar\partial_{p_n}+\alpha_nx_n\right),\ni\alpha_j+\gamma_j=1,\ j=1,\ldots,n
    \end{aligned}
\end{equation*}
\begin{equation*}
    \begin{aligned}
           H\left(x_1,\ldots,x_n;p_1,\ldots,p_n; t\right)\rightarrow\hat{H}\left({\hat{x} }_1,\ldots,{\hat{x}}_n;{\hat{p} }_1,\ldots,{\hat{p}}_n;t\right)
    \end{aligned}
\end{equation*}
\begin{equation}
\label{eq:2.1}
    \begin{aligned}
        \equiv \hat{H}\left(i\hbar\partial_{p_1}+\alpha_1x_1,\ldots,i\hbar\partial_{p_n}+ \alpha_nx_n;-i\hbar\partial_{x_1}+\gamma_1p_1,\ldots,-i\hbar\partial_{x_n}+\gamma_np_n;t\right).
    \end{aligned}
\end{equation}

For our particle anti-particle system, our frame of reference is the particle. The transformation to a spherical coordinate system and the separation of angular variables can be seen in \ref{section:AppendixA} where $l(l+1)$ is the angular coupling term, ${p_r}$ is the radial momentum, and $r$ is the radial distance from the particle to the anti-particle. For a 1D system, the Hamiltonian is:
\begin{equation}
    \begin{aligned}
        \hat{H}=\frac{{\hat{{p_r}}}^2}{2m}+\hat{V}(r).
    \end{aligned}
\end{equation}
For a 3D system, the Hamiltonian is:
\begin{equation}
\label{eq:2.3}
    \begin{aligned}
        \hat{H}=\sum^{N}{\frac{-i\hbar{\hat{p}}_r}{m\hat{r}}+\frac{{\hat{p}}_r^2}{2m}+\frac{\hbar^2l(l+1)}{2m{\hat{r}}^2}}+\hat{V}(r).
    \end{aligned}
\end{equation}
Typically, $\hat{r}=-i\hbar\frac{\partial}{\partial {p_r}}+\alpha r,\  \hat{p}_r=-i\hbar\frac{\partial}{\partial r}+\gamma {p_r}  $ by convention of the QPSR. The more general form presented in [3] is:
\begin{equation}
    \begin{aligned}
        \hat{r}=\alpha r+i\hbar\beta\frac{\partial}{\partial {p_r}},\ \hat{{p_r}}=\gamma {p_r}+i\hbar\delta\frac{\partial}{\partial r}\ ,
    \end{aligned}
\end{equation}
where $\hat{r},\hat{p}_r$ are bound by the condition $\beta\gamma-\alpha\delta=1$ in order to keep consistency of $\hat{r}$ and $\hat{p}_r$ with the Heisenberg uncertainty principle, $[\hat{r},\hat{p}_r]=1$ . For this paper we instead chose $\alpha=1$ and $\gamma=1$, such that $\beta-\delta=1$.
This convention will be used for the entirety of this paper.

\bigskip

\section{1D Time Independent Schrodinger Equation for Confined Ground State Quarkonium}
In this section we find solution of the time independent Schrodinger Equation for heavy quark anti-quark system in one dimension. For the confined nature of the system, the linear term of the Cornell potential is only considered. Thus, the Hamiltonian in phase space becomes
\begin{equation}
    \begin{aligned}
        \hat{H}=\frac{{\hat{{p_r}}}^2}{2m}+b\hat{r}.
    \end{aligned}
\end{equation}
Where $\hat{r}\equiv r+i\hbar\beta\frac{\partial}{\partial {p_r}},  \ \hat{{p_r}}\equiv {p_r}+i\hbar\delta\frac{\partial}{\partial r}$  via the QPSR. Our radial Schrodinger Time Independent Equation thus becomes
\begin{equation}
    \begin{aligned}
        \left[\frac{\left({p_r}+i\hbar\delta\frac{\partial}{\partial r}\right)^2}{2m}+b\left(r+i\hbar\beta\frac{\partial}{\partial {p_r}}\right)\right]\Psi\left(r,{p_r}\right)=E\Psi(r,{p_r}),
    \end{aligned}
\end{equation}
which equals
\begin{equation}
\label{eq:3.3}
    \begin{aligned}
        \frac{{p_r}^2\Psi\left(r,{p_r}\right)}{2m}+br\Psi\left(r,{p_r}\right)+ib\delta\hbar\frac{\partial \Psi\left(r,p\right)}{\partial {p_r}}-\frac{i{p_r}\beta\hbar}{m}\frac{\partial \Psi\left(r,{p_r}\right)}{\partial r}-\frac{\beta^2\hbar^2}{2m}\frac{\partial^2\Psi\left(r,{p_r}\right)}{\partial r^2}=E\Psi\left(r,{p_r}\right).
    \end{aligned}
\end{equation}
Making the substitution $A=\frac{{p_r}^2}{2m}+br$, we can formulate from the chain rule that,
\begin{equation}
    \begin{aligned}
         \frac{\partial\Psi\left(r,{p_r}\right)}{\partial r}\ &=\ \frac{\partial\Psi(A)}{\partial A}\frac{\partial A}{\partial r}\ =\ b\frac{\partial\Psi\left(A\right)}{\partial A},
        \\
        \frac{\partial\Psi\left(r,{p_r}\right)}{\partial {p_r}}\ &=\ \frac{\partial\Psi\left(A\right)}{\partial A}\frac{\partial A}{\partial {p_r}}\ =\ \frac{{p_r}}{m}\frac{\partial\Psi(A)}{\partial A},
        \\
        \frac{\partial^2\Psi(r,{p_r})}{\partial r^2}&=\frac{\partial^2\Psi\left(A\right)}{\partial A^2}\frac{\partial A^2}{\partial r^2}=b^2\frac{\partial^2\Psi\left(A\right)}{\partial A^2}.
    \end{aligned}
\end{equation}
This transforms \eqref{eq:3.3} into 
\begin{equation}
\label{eq:3.5}
    \begin{aligned}
        \frac{{p_r}^2\Psi\left(A\right)}{2m}+br\Psi\left(A\right)+\frac{ib\delta\hbar {p_r}}{m}\frac{\partial\Psi\left(A\right)}{\partial A}-\frac{i{p_r}\beta\hbar b}{m}\frac{\partial\Psi\left(A\right)}{\partial A}-\frac{\beta^2\hbar^2b^2}{2m}\frac{\partial^2\Psi\left(A\right)}{\partial A^2}=E\Psi\left(r,{p_r}\right).
    \end{aligned}
\end{equation}
As we recall, $\beta$ and $\delta$ are bound by the condition $\beta-\delta=1$, and from \eqref{eq:3.5} we see that by equating the coefficients of the first order terms,
\begin{equation}
    \begin{aligned}
        \ \frac{ib\delta\hbar {p_r}}{m}&=\frac{-i{p_r}\beta\hbar b}{m}, 
        \\
        \delta&=-\beta
    \end{aligned}
\end{equation}
must be satisfied for the first order terms to cancel each other out. With the condition $\beta-\delta=1$, we chose $\delta=-\frac{1}{2},\ \beta=\frac{1}{2}$. This will be used for the entirety of this paper, and we define our operators as
\begin{equation}
\label{eq:3.7}
    \begin{aligned}
        \hat{r}\equiv r+i\hbar\beta\frac{\partial}{\partial {p_r}},\ \hat{{p_r}}\equiv {p_r}+i\hbar\delta\frac{\partial}{\partial r}.
    \end{aligned}
\end{equation}
\eqref{eq:3.5} becomes
\begin{equation}
\label{eq:3.8}
    \begin{aligned}
        \ \frac{{p_r}^2\Psi\left(A\right)}{2m}+br\Psi\left(A\right)-\frac{\hbar^2b^2}{8m}\frac{\partial^2\Psi\left(A\right)}{\partial A^2}&=E\Psi\left(r,{p_r}\right),
        \\
        A\Psi\left(r,{p_r}\right)-\frac{\hbar^2b^2}{8m}\frac{\partial^2\Psi\left(A\right)}{\partial A^2}&=E\Psi\left(r,{p_r}\right),
        \\
        \omega\frac{\partial^2\Psi\left(A\right)}{\partial A^2}=-E\Psi\left(r,{p_r}\right)+A\Psi\left(r,{p_r}\right)&=\left(A-E\right)\Psi\left(r,{p_r}\right),
    \end{aligned}
\end{equation}
where $\omega= \frac{\hbar^2b^2}{8m}$. With substitution $\varepsilon=\frac{A-E}{w}$, we can represent \eqref{eq:3.8} as the Airy equation
\begin{equation}
    \begin{aligned}
        \frac{\partial^2\Psi\left(A\right)}{\partial A^2\ }-\varepsilon\Psi\left(r,{p_r}\right) = 0 
    \end{aligned}
\end{equation}
which yields the solution
\begin{equation}
\label{eq:3.10}
    \begin{aligned}
        \ \Psi\left(A\right)=c_1Ai\left[\frac{A-E}{\omega}\omega^\frac{2}{3}\ \right]+c_2Bi\left[\frac{A-E}{\omega}\omega^\frac{2}{3}\ \right].
    \end{aligned}
\end{equation}
where $Ai$ and $Bi$ are the special Airy function counterparts. \eqref{eq:3.10} can be simplified to
\begin{equation}
    \begin{aligned}
        \Psi\left(A\right)=c_1Ai\left[\left(A-E\right)\omega^{-\frac{1}{3}}\ \right]+c_2Bi\left[\left(A-E\right)\omega^{-\frac{1}{3}}\ \right],
    \end{aligned}
\end{equation}
And $A$ can be re-substituted to get $\Psi$ in terms of $r,{p_r}$:
\begin{equation}
    \begin{aligned}
        \Psi\left(r,{p_r}\right)=\ c_1Ai\left[\left(\frac{{p_r}^2}{2m}+br-E\right)\omega^{-\frac{1}{3}}\ \right]+c_2Bi\left[\left(\frac{{p_r}^2}{2m}+br-E\right)\omega^{-\frac{1}{3}}\ \right].
    \end{aligned}
\end{equation}
As $r$ approaches infinity, $Bi$ would also approach infinity. Thus, $c_2$ is zero for the wave function to converge: 
\begin{equation}
\label{eq:3.13}
    \begin{aligned}
        \Psi\left(r,{p_r}\right)=\ c_1Ai\left[\left(\frac{{p_r}^2}{2m}+br-E\right)\omega^{-\frac{1}{3}}\ \right].
    \end{aligned}
\end{equation}
The energy can be deduced by the condition $\Psi_n\left(0\right)=0$, resulting in 
\begin{equation}
    \begin{aligned}
        \left(\ \frac{{p_r}^2}{2m}+br-E_n\right)\omega^{-\frac{1}{3}}=z_{n+1}. 
    \end{aligned}
\end{equation}
Where $z_n$ is the nth zero of the Airy $Ai$ function starting from $n=1$. Then,
\begin{equation}
\label{eq:3.15}
    \begin{aligned}
        {E}_n=\frac{{p_r}^2}{2m}-z_{n+1}\left(\frac{b^2\hbar^2}{8m}\right)^\frac{1}{3}.
    \end{aligned}
\end{equation}
The wave function can be normalized for various kinetic energies while in the ground state. Since $r$ and ${p_r}$ are representations of spherical coordinates projected onto the 2D phase space plane, the normalization integral contains variable components over $(r,\theta,\phi)$ with a Jacobian matrix determinant scaling factor. Take note that the momentum's $\theta,\phi$ dependence is already fulfilled through the angular variable separation of the Schrodinger equation; thus, we only integrate across the radial momentum ${p_r}$. Our normalization condition becomes
\begin{equation}
    \begin{aligned}
        \int_{-\infty}^{\infty}\int_{0}^{2\pi}\int_{0}^{\pi}\int_{0}^{\infty}{\left|\Psi\left(r,{p_r}\right)\right|^2\ r^2\sin{\theta} \ drd\theta d\phi d{p_r}=1}.
    \end{aligned}
\end{equation}
Using \eqref{eq:3.15} in the normalization condition with ground state energy $E_0$ yields 
\begin{equation}
\label{eq:3.17}
    \begin{aligned}
        \int_{-\infty}^{\infty}\int_{0}^{2\pi}\int_{0}^{\pi}\int_{0}^{\infty}{\left|c_1Ai\left[\left(\frac{{p_r}^2}{2m}+br-E_0\right)\omega^{-\frac{1}{3}}\ \right]\right|^2\ r^2\sin{\theta} \ drd\theta d\phi d{p_r}}.
    \end{aligned}
\end{equation}
Making the substitution $\omega^\frac{-1}{3}=w,\  \rho=\left(-E_0+\frac{{p_r}^2}{2m}\right)w,\  q=b w r$, equation \eqref{eq:3.17} becomes
\begin{equation}
\label{eq:3.18}
    \begin{aligned}
        \ 2\int_{0}^{\infty}\int_{0}^{2\pi}\int_{0}^{\pi}\int_{0}^{\infty}{{\left|c_1\right|^2\left|Ai\left[\rho+q\right]\right|}^2\ \left(\frac{q}{bw}\right)^2\left(\frac{1}{bw}\right)\sin{\theta} \ dqd\theta d\phi d{p_r}}.
    \end{aligned}
\end{equation}
[7] establishes the relationship
\begin{equation}
\label{eq:3.19}
    \begin{aligned}
        \int_{0}^{\infty}{t^nAi^2\left[t+x\right]dt\ =\ \frac{n}{\left(2n+1\right)}\left[\frac{1}{2}\ \ \frac{d^{2}}{dt^2}-2t \right]\int_{0}^{\infty}{t^{n-1}Ai^2\left[t+x\right]dt,\ \ n>0}}.
    \end{aligned}
\end{equation}
Considering that the Airy $Ai$ function in the kernel of \eqref{eq:3.18} has strictly real values, applying \eqref{eq:3.19} to \eqref{eq:3.18} yields
\begin{equation}
\label{eq:3.20}
    \begin{aligned}
        2\int_{-\infty}^{\infty}\int_{0}^{2\pi}\int_{0}^{\pi}{\ \frac{2}{5}\left[\frac{1}{2}\frac{d^2}{d\rho^2}-2\rho\right]\int_{0}^{\infty}{{\left|c_1\right|^2\left(Ai\left[\rho+q\right]\right)}^2\frac{q}{\left(bw\right)^3}\sin{\theta} \ dqd\theta d\phi d\rho}}
    \end{aligned}
\end{equation}
Applying \eqref{eq:3.19} again to \eqref{eq:3.20} results in
\begin{equation*}
    \begin{aligned}
        2\int_{0}^{\infty}\int_{0}^{2\pi}\int_{0}^{\pi}{\frac{1}{3}\left[\frac{1}{2}\frac{d^2}{d\rho^2}-2\rho\right]\ \frac{2}{5}\left[\frac{1}{2}\frac{d^2}{d\rho^2}-2\rho\right]\int_{0}^{\infty}{{\left|c_1\right|^2\left(Ai\left[\rho+q\right]\right)}^2\frac{1}{\left(bw\right)^3}\sin{\theta} \ dqd\theta d\phi d\rho}}
    \end{aligned}
\end{equation*}
\begin{equation}
    \begin{aligned}
        &=2\left|c_1\right|^2\frac{1}{\left(bw\right)^3}\int_{0}^{\infty}\int_{0}^{2\pi}\int_{0}^{\pi}{\frac{1}{3}\left[\frac{1}{2}\frac{d^2}{d\rho^2}-2\rho\right]\ \frac{2}{5}\left[\frac{1}{2}\frac{d^2}{d\rho^2}-2\rho\right]\left(\frac{-\rho A i^2\left[\rho\right]+A{i^\prime}^2\left[\rho\right]}{b^3w^3}\right)\sin{\theta}d\theta d\phi d\rho} \\
        &=2\left|c_1\right|^2\frac{1}{\left(bw\right)^3}\int_{0}^{\infty}\int_{0}^{2\pi}\int_{0}^{\pi}{\left(\frac{\left(3-8\rho^3\right)Ai^2\left[\rho\right]+4\rho Ai\left[\rho\right]Ai^\prime\left[\rho\right] +8\rho^2A{i^\prime}^2\left[\rho\right]}{2b^3w^3}\right)\sin{\theta}d\theta d\phi d\rho} \\
        &=2\left|c_1\right|^2\frac{2\pi}{\left(bw\right)^3}\int_{0}^{\infty}\left(\frac{\left(3-8\rho^3\right)Ai^2\left[\rho\right]+4\rho Ai\left[\rho\right]Ai^\prime\left[\rho\right]+8\rho^2A{i^\prime}^2\left[\rho\right]}{2b^3w^3}\right)d\rho \\
        &=\frac{\left|c_1\right|^2\left(5\times3^\frac{1}{3}\right)}{6b^3w^3{\ \Gamma}^2\left(\frac{1}{3}\right)}.
    \end{aligned}
\end{equation}
Thus, our normalization constant is:
\begin{equation}
\label{eq:3.23}
    \begin{aligned}
        c_1=\sqrt{\frac{6b^3w^3{\ \Gamma}^2\left(\frac{1}{3}\right)}{\left(5\times3^\frac{1}{3}\right)}}.
    \end{aligned}
\end{equation}
Back-substituting w, \eqref{eq:3.23} resolves to
\begin{equation}
\label{eq:3.24}
    \begin{aligned}
        c_1=\sqrt{\frac{6b^3\left(\frac{\hbar^2b^2}{8m}\right)^3{\ \Gamma}^2\left(\frac{1}{3}\right)}{\left(5\times3^\frac{1}{3}\right)}}\ =\frac{7b^9\hbar^6{\ \Gamma}^2\left(\frac{1}{3}\right)}{2560\times3^\frac{1}{3}m^3}.
    \end{aligned}
\end{equation}
The wave function normalization constant, $c_1$, can be calculated from \eqref{eq:3.24} using the meson specific quantities m, the mass, and b, the linear term in the Cornell potential.
\\
\bigskip

\section{Nikiforov-Uvarov Method}
The Nikiforov-Uvarov method is based on solving the hyper-geometric type second-order differential equation of form
\begin{equation}
\label{eq:4.1}
    \begin{aligned}
        {\ \Psi}^{\prime\prime}\left(s\right)+\frac{\widetilde{\tau}}{\sigma\left(s\right)}\Psi^\prime\left(s\right)+\frac{\ \widetilde{\sigma}\left(s\right)}{\sigma^2\left(s\right)}\Psi\left(s\right)=0
    \end{aligned}
\end{equation}
Where $\sigma(s)$ and $\widetilde{\sigma}\left(s\right)$ are at most, second degree polynomials, and $\widetilde{\tau}\left(s\right)$ is at most a first-degree polynomial. $\Psi\left(s\right)$ is a function of hyper-geometric type. The solution of \eqref{eq:4.1} takes the form 
\begin{equation}
\label{eq:4.2}
    \begin{aligned}
        \Psi\left(s\right)=\phi\left(s\right)y\left(s\right)
    \end{aligned}
\end{equation}
Substituting \eqref{eq:4.2} into \eqref{eq:4.1} yields
\begin{equation}
    \begin{aligned}
        \sigma\left(s\right)y^{\prime\prime}\left(s\right)+\tau\left(s\right)y^\prime\left(s\right)+\lambda\left(s\right)y\left(s\right)=0,
    \end{aligned}
\end{equation}
where $\phi\left(s\right)$ satisfies the following relation
\begin{equation}
\label{eq:4.4}
    \begin{aligned}
        \frac{\phi^\prime\left(s\right)}{\phi(s)}=\frac{\pi\left(s\right)}{\sigma(s)}.
    \end{aligned}
\end{equation}
and $y\left(s\right)$ is a hyper-geometric type function, whose polynomial solutions are obtained from Rodrigues' relation 
\begin{equation}
    \begin{aligned}
        \ y\left(s\right)=y_n\left(s\right)=\frac{B_n}{\rho\left(s\right)}\frac{d^n}{ds^n}\left[\sigma^n\left(s\right)\rho\left(s\right)\right].
    \end{aligned}
\end{equation}
where $B_n$ is the normalization constant and $\rho\left(s\right)$ is a weight function satisfying the equation 
\begin{equation}
\label{eq:4.6}
    \begin{aligned}
        \left[\sigma\left(s\right)\rho\left(s\right)\right]^\prime=\tau\left(s\right)\rho\left(s\right).
    \end{aligned}
\end{equation}
The function $\pi\left(s\right)$ is defined as
\begin{equation}
\label{eq:4.7}
    \begin{aligned}
        \pi\left(s\right)=\left(\frac{\sigma^\prime-\widetilde{\tau}}{2}\right)\pm\sqrt{\left(\frac{\sigma^\prime-\widetilde{\tau}}{2}\right)^2-\widetilde{\sigma}+K\sigma},
    \end{aligned}
\end{equation}
and $\lambda$ is defined as 
\begin{equation}
\label{eq:4.8}
    \begin{aligned}
        \lambda=K+\pi\prime.
    \end{aligned}
\end{equation}
The value of $K$ can be calculated under the condition that the square root in \eqref{eq:4.7} must be the square of a polynomial. Thus, the equation of eigenvalues can be given as:
\begin{equation}
\label{eq:4.9}
    \begin{aligned}
        \lambda=\lambda_n=-n\tau^\prime-\frac{n\left(n-1\right)}{2}\sigma^{\prime\prime},
    \end{aligned}
\end{equation}
where
\begin{equation}
\label{eq:4.10}
    \begin{aligned}
        \tau\left(s\right)=\widetilde{\tau}\left(s\right)+2\pi\left(s\right).
    \end{aligned}
\end{equation}

\medskip

\section{Full Cornell Potential 3D Phase Space Schrodinger Equation}
In this section, the 3D Schrodinger Equation in phase space will be constructed along with a scheme to find the wave function and eigenvalues using the Nikiforov-Uvarov method and integral transforms.

Using the general QPSR definitions from \eqref{eq:2.1}, the time independent wave equation becomes	
\begin{equation}
\label{eq:5.1}
    \begin{aligned}
        \hat{H}\binom{i\hbar\partial_{p_{r_1}}+\alpha_1x_1,\ldots,i\hbar\partial_{p_{r_n}}+\alpha_nx_n;}{-i\hbar\partial_{x_1}+\gamma_1p_{r_1},\ldots,-i\hbar\partial_{x_n}+\gamma_np_{r_n}}\psi\left(x_1,\ldots,x_n;p_{r_1},\ldots,p_{r_n}\right)=E_n\psi(x_1,\ldots,x_n;p_{r_1},\ldots,p_{r_n}). 
    \end{aligned}
\end{equation}
Applying the alternative to the convolution and multi-variable inverse transform (3A), the convolution of ${p_r}$ between (4A) and relation (5A) yields
\begin{equation*}
    \begin{aligned}
        \left[L_{\binom{\left({\partial_{p_{r_1}},\ldots,\partial_{p_{r_n}}}\right)}{\rightarrow\left(p_{r_1},\ldots.,p_{r_n}\right)}}^{-1}\left[\hat{H}\binom{i\hbar\partial_{p_{r_1}}+\alpha_1x_1,\ldots,i\hbar\partial_{p_{r_n}}+\alpha_nx_n;}{-i\hbar\partial_{x_1}+\gamma_1p_{r_1},\ldots,-i\hbar\partial_{x_n}+\gamma_np_{r_n}}\right]\right] \\
        \ast \ \ \ \ \ \psi\left(x_1,\ldots,x_n;p_{r_1},\ldots,p_{r_n}\right)\\ 
    \end{aligned}
    \begin{tikzpicture}[baseline=22pt]

        \hspace*{-145.5pt}
        \draw [decorate,decoration={brace,amplitude=5pt}](.6,0.0) -- (-.2,0.0);  
    \end{tikzpicture}
    \begin{tikzpicture}[baseline=33.5pt]
        \hspace*{-190.5pt}
        \node at (0,0) {$\left(p_{r_1},\ldots,p_{r_n}\right)$};
    \end{tikzpicture}
\end{equation*}
\begin{equation}
\label{eq:5.2}
    \begin{aligned}
        \equiv \hat{H}\binom{i\hbar\partial_{p_{r_1}}+\alpha_1x_1,\ldots,i\hbar\partial_{p_{r_n}}+\alpha_nx_n;}{-i\hbar\partial_{x_1}+\gamma_1p_{r_1},\ldots,-i\hbar\partial_{x_n}+\gamma_np_{r_n}}\psi\left(x_1,\ldots,x_n;p_{r_1},\ldots,p_{r_n}\right).                        
    \end{aligned}
\end{equation}
Applying \eqref{eq:5.2}-\eqref{eq:5.1} with the convolution identities in (3A), \eqref{eq:5.2} is transformed into
\begin{equation}
    \begin{aligned}
       L_{\binom{\left({p_{r_1},\ldots,p_{r_n}}\right)} {\rightarrow\left({\overline{p}_r}_1,\ldots.,{\overline{p}_r}_n\right)}}\ &\left[ \hat{H}\binom{i\hbar\partial_{p_{r_1}}+\alpha_1x_1,\ldots,i\hbar\partial_{p_{r_n}}+\alpha_nx_n;}{-i\hbar\partial_{x_1}+\gamma_1p_{r_1},\ldots,-i\hbar\partial_{x_n}+\gamma_np_{r_n}}\psi\left(x_1,\ldots,x_n;p_{r_1},\ldots,p_{r_n}\right)\right]
        \\ & \equiv\ L_{\binom{\left({p_{r_1},\ldots,{p_r}}_n\right)}{\rightarrow\left({\overline{p}_r}_1,\ldots.,{\overline{p}_r}_n\right)}}
        \begin{bmatrix}
        L_{\binom{\left({\partial_{p_{r_1}},\ldots,\partial}_{p_{r_n}}\right)}{\rightarrow\left(p_{r_1},\ldots.,p_{r_n}\right)}}^{-1}\left[\hat{H}\binom{i\hbar\partial_{p_{r_1}}+\alpha_1x_1,\ldots,i\hbar\partial_{p_{r_n}}+\alpha_nx_n;}{-i\hbar\partial_{x_1}+\gamma_1p_{r_1},\ldots,-i\hbar\partial_{x_n}+\gamma_np_{r_n}}\right] \\
        \ast \ \ \ \ \ \ \ \ \psi\left(x_1,\ldots,x_n;p_{r_1},\ldots,p_{r_n}\right)
        \end{bmatrix}
        \\
        &=E_n\check{\psi}\left(x_1,\ldots,x_n;{\overline{p}_r}_1,\ldots,{\overline{p}_r}_n\right)
    \end{aligned}
    \begin{tikzpicture}[baseline=23.5pt]

        \hspace*{-200pt}
        \draw [decorate,decoration={brace,amplitude=5pt}](.6,0.0) -- (-.2,0.0);  
    \end{tikzpicture}
    \begin{tikzpicture}[baseline=30pt]
        \hspace*{-242.5pt}
        \node at (0,0) {$\left(p_{r_1},\ldots,p_{r_n}\right)$};
    \end{tikzpicture}
\end{equation}
\begin{equation}
    \begin{aligned}
        \equiv\hat{H}\binom{i\hbar{\overline{p}_r}_1+\alpha_1x_1,\ldots,i\hbar{\overline{p}_r}_n+\alpha_nx_n;}{-i\hbar\partial_{x_1}+\gamma_1p_{r_1},\ldots,-i\hbar\partial_{ x_n}+\gamma_np_{r_n}}\check{\psi}\left(x_1,\ldots,x_n;{\overline{p}_r}_1,\ldots,{\overline{p}_r}_n\right) ={E}_n\check{\psi}\left(x_1,\ldots,x_n;{\overline{p}_r}_1,\ldots,{\overline{p}_r}_n\right).
    \end{aligned}
\end{equation}
Thus, the wave function in phase space can be expressed as the inverse transform of the solution $\breve{\psi}\left(x_1,\ldots,x_n;{\overline{p}_r}_1,\ldots,{\overline{p}_r}_n;t\right)$. The radial Schrodinger equation using the 3D full Cornell potential Hamiltonian is
\begin{equation}
    \begin{aligned}
        \left(-\frac{i\hbar\hat{{p_r}}}{m\hat{r}}+\frac{{\hat{{p_r}}}^2}{2m}+\frac{\hbar^2l\left(l+1\right)}{2m\hat{r}}+\frac{a}{\hat{r}}+b\hat{r}\right)\psi\left(r,{p_r}\right)=E_n\psi\left(r,\overline{p}_r\right).
    \end{aligned}
\end{equation}
It is worth noting that ${p_r}$, once again, only describes the radial momentum after the angular separation of the Schrodinger equation. Using the specific operator definitions in \eqref{eq:3.7} with the symmetrical coefficients $\delta=-\frac{1}{2}, \beta=\frac{1}{2}$ and mapping $(\partial_{{p_r}}\rightarrow\overline{p}_r)$:
\begin{equation}
    \begin{aligned}
        \begin{pmatrix}
        &\frac{-i\left(-\frac{i\hbar}{2}\frac{\partial}{\partial r}+{p_r}\right)}{m\left(\frac{i\hbar}{2}\overline{p}_r+r\right)}+\frac{\left(-\frac{i\hbar}{2}\frac{\partial}{\partial r}+{p_r}\right)^2}{2m}+\frac{\hbar^2l\left(l+1\right)}{2m\left(\frac{i\hbar}{2}\overline{p}_r+r\right)^2} \\
        &+\frac{a}{\left(\frac{i\hbar}{2}\overline{p}_r+r\right)}+b\left(\frac{i\hbar}{2}\overline{p}_r+r\right)
        \end{pmatrix}\psi\left(r,\overline{p}_r\right)=E_n\psi(r,\overline{p}_r)
    \end{aligned}
\end{equation}
\begin{equation}
\label{eq:5.7}
    \begin{aligned}
        \begin{pmatrix}
        \frac{l\left(1+l\right)\hbar^2\ \ \psi\left(r,\overline{p}_r\right)}{2m\left(\frac{i\hbar}{2}\overline{p}_r+r\right)^2}+\frac{a\ \psi\left(r,\overline{p}_r\right)}{r+\frac{i\hbar\overline{p}_r}{2}}+b\left(r+\frac{i\hbar\overline{p}_r}{2}\right)\psi\left(r,\overline{p}_r\right) \\
        -\frac{i\hbar\left({p_r}\ \psi\left(r,\overline{p}_r\right)-\frac{i\hbar}{2}\frac{\partial\psi\left(r,\overline{p}_r\right)\ }{\ \partial r}\right)\ }{m\left(r+\frac{i\hbar\overline{p}_r}{2}\right)}+\frac{{p_r}\left({p_r}\ \psi\left(r,\overline{p}_r\right)-\frac{i\hbar}{2}\frac{\partial\psi\left(r,\overline{p}_r\right)}{\partial r}\right)-\frac{i\hbar}{2}\left({p_r}\frac{\partial\psi\left(r,\overline{p}_r\right)}{\partial r}-\frac{i\hbar}{2}\frac{\partial^2\psi\left(r,\overline{p}_r\right)}{\partial r^2}\right)\ }{2m}
        \end{pmatrix}=E_n\psi\left(r,\overline{p}_r\right).
    \end{aligned}
\end{equation}
To generate a single variable function from a collective variable involving $r$ and ${p_r}$ with constant partial derivatives, as was done with the 1D quark confinement scenario, the substitution $A=r+\frac{i\hbar\overline{p}_r}{2}$ is made. The chain rule yields
\begin{equation}
    \begin{aligned}
        \ \frac{\partial\psi\left(r,\overline{p}_r\right)}{\partial r}=\frac{\partial\psi(A)}{\partial\ A}\frac{\partial\ A}{\partial r}=\frac{\partial\psi\left(A\right)}{\partial r},\ \ \frac{\partial^2\psi\left(r,\overline{p}_r\right)}{\partial r^2}=\frac{\partial^2\psi(A)}{\partial\ A^2}\frac{\partial\ A^2}{\partial r^2}=\frac{\partial^2\psi\left(A\right)}{\partial r^2},
    \end{aligned}
\end{equation}
a direct mapping from $\left(r,\overline{p}_r\right)$ to $A$. \eqref{eq:5.7} transforms into
\begin{equation}
    \begin{aligned}
        \begin{pmatrix}
            \left(\frac{a}{A}+Ab-E_n+\frac{{p_r}^2}{2m}-\frac{i{p_r}\hbar}{Am}+\frac{l\left(l+1\right)}{2A^2m}\right)\psi\left(A\right)-\frac{i{p_r}\hbar\ \psi\prime\left(A\right)}{4m}-\frac{\hbar^2\psi^\prime\left(A\right)}{2mA} \\
            -\frac{i\hbar\left({p_r}\ \psi^\prime\left(A\right)-\frac{1}{2}i\hbar\ \psi^{\prime\prime}\left(A\right)\right)}{4m}
        \end{pmatrix}=0
    \end{aligned}
\end{equation}
\begin{equation}
\label{eq:5.10}
    \begin{aligned}
         =\left(\frac{a}{A}+Ab-E_n+\frac{{p_r}^2}{2m}-\frac{i{p_r}\hbar}{Am}+\frac{l\left(l+1\right)}{2A^2m}\right)\psi\left(A\right)-\frac{\left(iA{p_r}\hbar+\hbar^2\right)\ \psi^\prime\left(A\right)}{2Am}-\frac{\hbar^2\psi^{\prime\prime}\left(A\right)}{8m}=0.
    \end{aligned}
\end{equation}
Making the substitution $x=\frac{1}{A}$, the chain rule yields:
\begin{equation}
    \begin{aligned}
        \frac{d\psi(A)}{dA}=\frac{d\psi(x)}{dx}\frac{dx}{dA}=\ -\frac{1}{A^2}\frac{d\psi\left(x\right)}{dx},\ \ \frac{d^2\psi\left(A\right)}{dA^2}=\frac{d^2\psi\left(x\right)}{dx^2}\frac{dx^2}{dA^2}=\frac{1}{A^4}\frac{d^2\psi\left(x\right)}{dx^2}.
    \end{aligned}
\end{equation}
Then, \eqref{eq:5.10} becomes
\begin{equation}
\label{eq:5.12}
    \begin{aligned}
        \left(-E_n+\frac{{p_r}^2}{2m}+\frac{b}{x}+ax-\frac{i{p_r}x\hbar}{m}+\frac{l\left(l+1\right)x^2\hbar^2}{2m}\right)\psi\left(x\right)-\frac{x^2\left(i{p_r}\hbar+x\hbar^2\right)\psi^\prime\left(x\right)}{2m}- \frac{{x^4\hbar}^2\psi^{\prime\prime}\left(x\right)}{8m}=0 
    \end{aligned}
\end{equation}
The $\left(\frac{1}{x}\right)$ term can be approximated by assuming there is a characteristic radius in which the quark and anti-quark can interact with one another without colliding, denoted $r_0$. We also assume there is a variable ${\overline{p}_r}_0$, the value of the radial momentum in the transformed phase space the anti-quark obtains when at $r_0$. Since $x=\frac{1}{A}$, the $\left(\frac{1}{x}\right)$ term will be centered in the x space around $\frac{1}{r_0+\frac{i\hbar}{2}{\overline{p}_r}_0}=\frac{1}{A_0}=\delta$. We let $y=x-\delta$ around the singularity $y=0$, generating the power series to second degree: 
\begin{equation}
\label{eq:5.13}
    \begin{aligned}
        \ \left(\frac{1}{x}\right)=\frac{1}{\left(y+\delta\right)}=\left(\frac{1}{\delta}\right)\left(\frac{1}{1+\frac{y}{\delta}}\right)=\ \frac{3}{\delta}-\frac{3x}{\delta^2}+\frac{x^2}{\delta^3}.
    \end{aligned}
\end{equation}
Substituting \eqref{eq:5.13} into \eqref{eq:5.12} we get
\begin{equation}
    \begin{aligned}
        \begin{pmatrix}
            \left(-E_n+\frac{{p_r}^2}{2m}+b\left(\ \frac{3}{\delta}-\frac{3x}{\delta^2}+\frac{x^2}{\delta^3}\right)+ax-\frac{i{p_r}x\hbar}{m}+\frac{l\left(l+1\right)x^2\hbar^2}{2m}\right)\psi\left(x\right) \\
            -\frac{x^2\left(i{p_r}\hbar+x\hbar^2\right)\psi^\prime\left(x\right)}{2m}-\frac{{x^4\hbar}^2\psi^{\prime\prime}\left(x\right)}{8m}
        \end{pmatrix}=0,
    \end{aligned}
\end{equation}
which simplifies to 
\begin{equation}
    \begin{aligned}
        \begin{pmatrix}
            \left(-E_n+\frac{{p_r}^2}{2m}+\frac{3b}{\delta}-\frac{3bx}{\delta^2}+\frac{bx^2}{\delta^3}+ax-\frac{i{p_r}x\hbar}{m}+\frac{l\left(l+1\right)x^2\hbar^2}{2m}\right)\psi\left(x\right) \\
            -\frac{x^2\left(i{p_r}\hbar+x\hbar^2\right)\psi^\prime\left(x\right)}{2m}-\frac{{x^4\hbar}^2\psi^{\prime\prime}\left(x\right)}{8m}.
        \end{pmatrix}=0
    \end{aligned}
\end{equation}
\begin{equation}
    \begin{aligned}
        \begin{pmatrix}
            \ \ \ \ \frac{\left(-E_n+\frac{{p_r}^2}{2m}+\frac{3b}{\delta}-\frac{3bx}{\delta^2}+\frac{bx^2}{\delta^3}+ax-\frac{i{p_r}x\hbar}{m}+\frac{l\left(l+1\right)x^2\hbar^2}{2m}\right)8m}{\hbar^2}\psi\left(x\right) \\
            -\frac{4x^2\left(i{p_r}\hbar+x\hbar^2\right)\psi^\prime\left(x\right)}{\hbar^2}-x^4\psi^{\prime\prime}\left(x\right)
        \end{pmatrix}=0
    \end{aligned}
\end{equation}
\begin{equation}
\label{eq:5.17}
    \begin{aligned}
        \begin{pmatrix}
            \left(x^2\left(4l\left(l+1\right)+\frac{8bm}{\delta^3\hbar^2}\right)+x\left(\frac{8am}{\hbar^2}-\frac{24bm}{\delta^2\hbar^2}-\frac{8i{p_r}}{\hbar}\right)+\left(-\frac{8E_nm}{\hbar^2}+\frac{24bm}{\delta\hbar^2}+\frac{4{p_r}^2}{\hbar^2}\right)\right)\psi\left(x\right) \\
            -\frac{4x^2\left(i{p_r}\hbar+x\hbar^2\right)\psi^\prime\left(x\right)}{\hbar^2}-x^4\psi^{\prime\prime}\left(x\right)
        \end{pmatrix}=0
    \end{aligned}
\end{equation}
For convenience, we define three constants, $\alpha,\beta,\gamma$, to be 
\begin{equation}
    \begin{aligned}
        \alpha\equiv-\frac{8E_nm}{\hbar^2}+\frac{4{p_r}^2}{\hbar^2}+\frac{24bm}{\delta\hbar^2},\ \ \beta\equiv\frac{8am}{\hbar^2}-\frac{24bm}{\delta^2\hbar^2}-\frac{8i{p_r}}{\hbar},\ \ \gamma\equiv4L(1+L)+\frac{8bm}{\delta^3\hbar^2}.
    \end{aligned}
\end{equation}
Now, we can rewrite \eqref{eq:5.17} as
\begin{equation}
    \begin{aligned}
        \left(\alpha+x\beta+x^2\gamma\right)\psi\left(x\right)-\frac{4x^2\left(i{p_r}\hbar+x\hbar^2\right)\psi^\prime\left(x\right)}{\hbar^2}-x^4\psi^{\prime\prime}\left(x\right)=0,
    \end{aligned}
\end{equation}
which can be further simplified to 
\begin{equation}
\label{eq:5.20}
    \begin{aligned}
        \frac{\left(-\alpha-x\beta-x^2\gamma\right)}{x^4}\ \psi\left(x\right)+\frac{4\left(x+\frac{i{p_r}}{\hbar}\right)}{x^2}{\ \psi}^\prime\left(x\right)+{\ \psi}^{\prime\prime}\left(x\right)=0.
    \end{aligned}
\end{equation}
The exact form as in \eqref{eq:4.1} to apply the Nikiforov-Uvarov method. Thus,
\begin{equation}
    \begin{aligned}
        \sigma\left(x\right)=x^2, \quad \widetilde{\sigma}\left(x\right)=-\alpha-x\beta-x^2\gamma, \quad
        \widetilde{\tau}\left(x\right)=4\left(x+\frac{i{p_r}}.{\hbar}\right)
    \end{aligned}
\end{equation}
From \eqref{eq:4.7}
\begin{equation}
    \begin{aligned}
        \pi\left(x\right)=\frac{1}{2}\left(2x-4\left(x+\frac{i{p_r}}{\hbar}\right)\right)\pm\sqrt{Kx^2+\alpha+x\beta+x^2\gamma+\frac{1}{4}\left(2x-4\left(x+\frac{i{p_r}}{\hbar}\right)\right)^2}\ 
    \end{aligned}
\end{equation}
Since the polynomial under the radical must a square of a polynomial, the discriminant must equal zero.
\begin{equation}
    \begin{aligned}
        \ \left(\beta+\frac{4i{p_r}}{\hbar}\right)^2-4\left(\alpha-\frac{4{p_r}^2}{\hbar^2}\right)\left(1+K+\gamma\right)=0
    \end{aligned}
\end{equation}
Solving for $K$,
\begin{equation}
    \begin{aligned}
        K=\frac{-16{p_r}^2\gamma-8i{p_r}\beta\hbar+4\alpha\hbar^2-\beta^2\hbar^2+4\alpha\gamma\hbar^2}{4\left(4{p_r}^2-\alpha\hbar^2\right)}.
    \end{aligned}
\end{equation}
Thus,
\begin{equation}
\label{eq:5.25}
    \begin{aligned}
        \pi\left(x\right)=-x-\frac{2i{p_r}}{\hbar}\pm\frac{1}{2}\frac{\left(-8{p_r}^2+4i{p_r}x\hbar+\left(2\alpha+x\beta\right)\hbar^2\right)}{\sqrt{-4{p_r}^2\hbar^2+\alpha\hbar^4}}
    \end{aligned}.
\end{equation}
The choice of using the plus sign in \eqref{eq:5.25} will be used in defining $\pi(x)$ to generate a negative first derivative in $\tau$. From \eqref{eq:4.4}, $\phi(x)$ can be found by the differential equation
\begin{equation}
    \begin{aligned}
        \frac{\phi^\prime\left(x\right)}{\phi(x)}=\frac{\pi\left(x\right)}{\sigma(x)}=\frac{-x-\frac{2i{p_r}}{\hbar}+\frac{1}{2}\frac{\left(-8{p_r}^2+4i{p_r}x\hbar+\left(2\alpha+x\beta\right)\hbar^2\right)}{\sqrt{-4{p_r}^2\hbar^2+\alpha\hbar^4}}}{x^2},
    \end{aligned}
\end{equation}
which yields the exponential solution
\begin{equation}
    \begin{aligned}
        \phi\left(x\right)=e^{\left(\frac{8{p_r}^2-2\alpha\hbar^2+4i{p_r}\sqrt{-4{p_r}^2+\alpha\hbar^2}+x\hbar\left(4i{p_r}x-\left(2\sqrt{-4{p_r}^2+\alpha\hbar^2}\right)\right)\ln{\left(x\right)}}{2x\hbar\sqrt{-4{p_r}^2+\alpha\hbar^2}}\right)}.
    \end{aligned}
\end{equation}
By \eqref{eq:4.10}, $\tau$ is 
\begin{equation}
    \begin{aligned}
        \tau\left(x\right)=\widetilde{\tau}\left(x\right)+2\pi\left(x\right)=\frac{-8{p_r}^2+4i{p_r}x\hbar+\hbar\left(2\alpha\hbar+x\beta\hbar+2x\sqrt{-4{p_r}^2\alpha\hbar^2}\right)}{\hbar\sqrt{-4{p_r}^2+\alpha\hbar^2}}.
    \end{aligned}
\end{equation}
\eqref{eq:4.6} is used to find the weight function $\rho\left(x\right)$, where 
\begin{equation}
    \begin{aligned}
        \left[\sigma\left(x\right)\rho\left(x\right)\right]^\prime=\tau\left(x\right)\rho\left(x\right),
    \end{aligned}
\end{equation}
\begin{equation}
    \begin{aligned}
        2x\rho\left(x\right)+x^2\rho^\prime\left(x\right)=\left(\frac{-8{p_r}^2+4i{p_r}x\hbar+\hbar\left(2\alpha\hbar+x\beta\hbar+2x\sqrt{-4{p_r}^2\alpha\hbar^2}\right)}{\hbar\sqrt{-4{p_r}^2+\alpha\hbar^2}}\right)\rho\left(x\right),
    \end{aligned}
\end{equation}
which gives the solution
\begin{equation}
    \begin{aligned}
        \rho\left(x\right)=e^\frac{8{p_r}^2-2\alpha\hbar^2+x\hbar\left(4i{p_r}+\beta\hbar\right)\ln{x}}{x\hbar\sqrt{-4{p_r}^2+\alpha\hbar^2}}.
    \end{aligned}
\end{equation}
With $\rho(x)$ and $\sigma\left(x\right)$, the function $y$ can be formulated as
\begin{equation}
    \begin{aligned}
        y_n\left(x\right)=\frac{B}{\rho\left(x\right)}\frac{d^n}{dx^n}\left(\sigma^n\left(x\right)\rho\left(x\right)\right).
    \end{aligned}
\end{equation}
Where $B$ is a normalization constant. Recalling \eqref{eq:4.2}, the wave function must be 
\begin{equation}
\label{eq:5.33}
    \begin{aligned}
        \psi_n\left(x\right)=y_n\left(x\right)\phi\left(x\right)=\phi\left(x\right)\frac{B}{\rho\left(x\right)}\ \frac{d^n}{dx^n}\left(\sigma^n\left(x\right)\rho\left(x\right)\right)
    \end{aligned}
\end{equation}
Back-substituting $x\rightarrow\frac{1}{A}=\frac{1}{r+\frac{i\hbar}{2}\overline{p}_r}$ , the term $\phi\left(x\right)\frac{B}{\rho\left(x\right)}$ is
\begin{equation}
\label{eq:5.34}
    \begin{aligned}
        \phi(x)\frac{B}{\rho\left(x\right)}&=B \times{}Exp\begin{pmatrix}
            -\frac{\left(\frac{r+i\hbar\overline{p}_r}{2}\right)\left(8{p_r}^2-2\alpha\hbar^2+\frac{\hbar\left(4i{p_r}+\beta\hbar\right)\ln{\left(\frac{1}{r+\frac{i\hbar\overline{p}_r}{2}}\right)}}{r+\frac{i\hbar\overline{p}_r}{2}}\right)}{\hbar\sqrt{-4{p_r}^2+\alpha\hbar^2}} \\
            +\ \frac{\left(r+\frac{i\hbar\overline{p}_r}{2}\right)\left(8{p_r}^2-2\alpha\hbar^2+4i{p_r}\sqrt{-4{p_r}^2+\alpha\hbar^2}+\left(\frac{\hbar\left(4i{p_r}+\beta\hbar-2\sqrt{-4{p_r}^2+\alpha\hbar^2}\right)}{r+\frac{i\hbar\overline{p}_r}{2}}\right)\ln{\left(\frac{1}{r+\frac{i\overline{p}_r\hbar}{2}}\right)}\right)}{2\hbar\sqrt{-4{p_r}^2+\alpha\hbar^2}} 
        \end{pmatrix} \\
        &=Be^\frac{\left(2r+i\hbar\overline{p}_r\right)\left(2i{p_r}+\sqrt{-4{p_r}^2+\alpha\hbar^2}\right)}{2\hbar}\left(\frac{1}{r+\frac{i\hbar\overline{p}_r}{2}}\right)^{-1-\frac{4i{p_r}+\beta\hbar}{2\sqrt{-4{p_r}^2+\alpha\hbar^2}}}.
    \end{aligned}
\end{equation}
The term $\frac{d^n}{dx^n}\left(\sigma^n\left(x\right)\rho\left(x\right)\right) $ can also undergo the substitution $x\rightarrow\frac{1}{A}=\frac{1}{r+\frac{i\hbar}{2}\overline{p}_r}$ to yield:
\begin{equation}
\label{eq:5.35}
    \begin{aligned}
        \ \frac{d^n}{dx^n}\left(\sigma^n\left(x\right)\rho\left(x\right)\right)&=\frac{d^n}{d\left(\frac{1}{r+\frac{i\hbar\overline{p}_r}{2}}\right)^n}\left(\sigma^n\left(\frac{1}{r+\frac{i\hbar\overline{p}_r}{2}}\right)\rho\left(\frac{1}{r+\frac{i\hbar\overline{p}_r}{2}}\right)\right) \\
        &=\frac{d^n}{d\left(\frac{1}{r+\frac{i\hbar\overline{p}_r}{2}}\right)^n}\left(\left(\frac{1}{r+\frac{i\hbar\overline{p}_r}{2}}\right)^2e^\frac{\left(r+\frac{i\hbar\overline{p}_r}{2}\right)\left(8{p_r}^2-2\alpha\hbar^2+\frac{\hbar\left(4i{p_r}+\beta\hbar\right)\ln{\left(\frac{1}{r+\frac{i\hbar\overline{p}_r}{2}}\right)}}{r+\frac{i\hbar\overline{p}_r}{2}}\right)}{\hbar\sqrt{-4{p_r}^2+\alpha\hbar^2}}\right)
    \end{aligned}
\end{equation}
Substituting \eqref{eq:5.35} and \eqref{eq:5.34} back into \eqref{eq:5.33}:
\begin{equation}
    \begin{aligned}
        \ \psi_n\left(r,\overline{p}_r\right)&= \\
        &\begin{pmatrix}
            Be^\frac{\left(2r+i\hbar\overline{p}_r\right)\left(2i{p_r}+\sqrt{-4{p_r}^2+\alpha\hbar^2}\right)}{2\hbar}\left(\frac{1}{r+\frac{i\hbar\overline{p}_r}{2}}\right)^{-1-\frac{4i{p_r}+\beta\hbar}{2\sqrt{-4{p_r}^2+\alpha\hbar^2}}} \\
            \times 
            \frac{d^n}{d\left(\frac{1}{r+\frac{i\hbar\overline{p}_r}{2}}\right)^n}\left(\left(\frac{1}{r+\frac{i\hbar\overline{p}_r}{2}}\right)^2e^\frac{\left(r+\frac{i\hbar\overline{p}_r}{2}\right)\left(8{p_r}^2-2\alpha\hbar^2+\frac{\hbar\left(4i{p_r}+\beta\hbar\right)\ln{\left(\frac{1}{r+\frac{i\hbar\overline{p}_r}{2}}\right)}}{r+\frac{i\hbar\overline{p}_r}{2}}\right)}{\hbar\sqrt{-4{p_r}^2+\alpha\hbar^2}}\right)
        \end{pmatrix}.
    \end{aligned}
\end{equation}
Thus, the wave function in $r,{p_r}$ is 
\begin{equation}
\label{eq:wavefunction}
    \begin{aligned}
        \psi_n\left(r,{p_r}\right) &= \\
            &L_{\left(\overline{p}_r\rightarrow\ {p_r}\right)}^{-1}\begin{pmatrix}
               Be^\frac{\left(2r+i\hbar\overline{p}_r\right)\left(2i{p_r}+\sqrt{-4{p_r}^2+\alpha\hbar^2}\right)}{2\hbar}\left(\frac{1}{r+\frac{i\hbar\overline{p}_r}{2}}\right)^{-1-\frac{4i{p_r}+\beta\hbar}{2\sqrt{-4{p_r}^2+\alpha\hbar^2}}} \\
               \times\frac{d^n}{d\left(\frac{1}{r+\frac{i\hbar\overline{p}_r}{2}}\right)^n}\left(\left(\frac{1}{r+\frac{i\hbar\overline{p}_r}{2}}\right)^2e^\frac{\left(r+\frac{i\hbar\overline{p}_r}{2}\right)\left(8{p_r}^2-2\alpha\hbar^2+\frac{\hbar\left(4i{p_r}+\beta\hbar\right)\ln{\left(\frac{1}{r+\frac{i\hbar\overline{p}_r}{2}}\right)}}{r+\frac{i\hbar\overline{p}_r}{2}}\right)}{\hbar\sqrt{-4{p_r}^2+\alpha\hbar^2}}\right).
            \end{pmatrix}
    \end{aligned}
\end{equation}
The corresponding energy eigenvalues are found via \eqref{eq:4.8} and \eqref{eq:4.9}.
\begin{equation}
    \begin{aligned}
        \lambda=K+\pi^\prime\left(x\right)=\ -1+\frac{-16{p_r}^2\gamma-8i{p_r}\beta\hbar+4\alpha\hbar^2-\beta^2\hbar^2+4\alpha\gamma\hbar^2}{4\left(4{p_r}^2-\alpha\hbar^2\right)}+\frac{4i{p_r}\hbar+\beta\hbar^2}{2\hbar\sqrt{-4{p_r}^2+\alpha\hbar^2}},
    \end{aligned}
\end{equation}
\begin{equation}
    \begin{aligned}
        \lambda_n=\ -n\tau^\prime\left(x\right)-\frac{n\left(n-1\right)\sigma^{\prime\prime}\left(x\right)}{2}=\frac{-n\left(4i{p_r}+\beta\hbar+\left(n+1\right)\sqrt{-4{p_r}^2+\alpha\hbar^2}\right)}{\sqrt{-4{p_r}^2+\alpha\hbar^2}}.
    \end{aligned}
\end{equation}
With condition $\lambda=\lambda_n$, 
\begin{equation}
\label{eq:5.40}
    \begin{aligned}
        -1&+\frac{-16{p_r}^2\gamma-8i{p_r}\beta\hbar+4\alpha\hbar^2-\beta^2\hbar^2+4\alpha\gamma\hbar^2}{4\left(4{p_r}^2-\alpha\hbar^2\right)}+\frac{4i{p_r}\hbar+\beta\hbar^2}{2\hbar\sqrt{-4{p_r}^2+\alpha\hbar^2}} \\
        &=\frac{-n\left(4i{p_r}+\beta\hbar+\left(n+1\right)\sqrt{-4{p_r}^2+\alpha\hbar^2}\right)}{\sqrt{-4{p_r}^2+\alpha\hbar^2}}.
    \end{aligned}
\end{equation}
$\alpha\equiv-\frac{8E_nm}{\hbar^2}+\frac{4{p_r}^2}{\hbar^2}+\frac{24bm}{\delta\hbar^2}$, so the substitution $\alpha\equiv cE_n+d$ is made, where 
\begin{equation}
    \begin{aligned}
        c=\ -\frac{8m}{\hbar^2},\quad d=\frac{4{p_r}^2}{\hbar^2}+\frac{24bm}{\delta\hbar^2}.
    \end{aligned}
\end{equation}
\eqref{eq:5.40} then becomes 
\begin{equation}
\label{eq:5.42}
    \begin{aligned}
        -1&+\frac{-16{p_r}^2\gamma-8i{p_r}\beta\hbar+4(cE_n+d)\hbar^2-\beta^2\hbar^2+4(cE_n+d)\gamma\hbar^2}{4\left(4{p_r}^2-\left(cE_n+d\right)\hbar^2\right)}+\frac{4i{p_r}\hbar+\beta\hbar^2}{2\hbar\sqrt{-4{p_r}^2+(cE_n+d)\hbar^2}} \\
        &=-\frac{n\left(4i{p_r}+\beta\hbar+\left(n+1\right)\sqrt{-4{p_r}^2+\alpha\hbar^2}\right)}{\sqrt{-4{p_r}^2+\alpha\hbar^2}}.
    \end{aligned}
\end{equation}
Solving \eqref{eq:5.42} for $E_n$ yields the result 
\begin{equation}
    \begin{aligned}
        E_n=\frac{\begin{pmatrix}
            \pm\sqrt{\left(c+2cn\right)^2\left(9+4\gamma\right)\hbar^4\left(4i{p_r}+\beta\hbar\right)^4} \\+c\hbar^2\begin{pmatrix}
                16{p_r}^2\left(3+6\gamma+2\left(n\left(1+n\right)\left(-5+n+n^2\right)-2n\left(n+1\right)\gamma+\gamma^2\ \right)\ \right) \\
                +\ 8i{p_r}\beta\left(5+2n\left(n+1\right)+2\gamma\right)\hbar \\
                +\ \left(-8d\left(-2+n+n^2-\gamma\right)^2+\beta^2\left(5+2n\left(n+1\right)+2\gamma\right)\right)\hbar^2
            \end{pmatrix}
        \end{pmatrix}}{8c^2\left(-2+n+n^2-\gamma\right)^2\hbar^4}.
    \end{aligned}
\end{equation}
Substituting the values of $c,d,\beta$, and $\gamma$ results in 
\begin{equation}
\label{eq:energyeigenvalues}
    \begin{aligned}
        E_n=& \\
        &\frac{\begin{pmatrix}
            \pm\sqrt{\left(-\frac{8m}{\hbar^2}-\frac{16m}{\hbar^2}n\right)^2\left(9+4\left(4L\left(1+L\right)+\frac{8bm}{\delta^3\hbar^2}\right)\right)\hbar^4\left(4i{p_r}+\left(\frac{8am}{\hbar^2}-\frac{24bm}{\delta^2\hbar^2}-\frac{8i{p_r}}{\hbar}\right)\hbar\right)^4} \\
            -\ \frac{8m}{\hbar^2}\hbar^2\begin{pmatrix}
                16{p_r}^2\begin{pmatrix}
                    3+6\left(4L\left(1+L\right)+\frac{8bm}{\delta^3\hbar^2}\right) \\
                    +2\begin{pmatrix}
                       n\left(1+n\right)\left(-5+n+n^2\right)-2n\left(n+1\right)\left(4L\left(1+L\right)+\frac{8bm}{\delta^3\hbar^2}\right) \\
                       +\left(4L\left(1+L\right)+\frac{8bm}{\delta^3\hbar^2}\right)^2
                    \end{pmatrix}
                \end{pmatrix}
                \\
                +\ 8i{p_r}\left(\frac{8am}{\hbar^2}-\frac{24bm}{\delta^2\hbar^2}-\frac{8i{p_r}}{\hbar}\right)\left(5+2n\left(n+1\right)+2\left(4L\left(1+L\right)+\frac{8bm}{\delta^3\hbar^2}\right)\right)\hbar 
                \\
                +\hbar^2\begin{pmatrix}
                    -8\left(\frac{4{p_r}^2}{\hbar^2}+\frac{24bm}{\delta\hbar^2}\right)\left(-2+n+n^2-\left(4L\left(1+L\right)+\frac{8bm}{\delta^3\hbar^2}\right)\right)^2 \\
                    +\left(\frac{8am}{\hbar^2}-\frac{24bm}{\delta^2\hbar^2}-\frac{8i{p_r}}{\hbar}\right)^2\left(5+2n\left(n+1\right)+2\left(4L(1+L)+\frac{8bm}{\delta^3\hbar^2}\right)\right)
                \end{pmatrix}
            \end{pmatrix}
        \end{pmatrix}}{8\left(-\frac{8m}{\hbar^2}\right)^2\left(-2+n+n^2-\left(4L(1+L)+\frac{8bm}{\delta^3\hbar^2}\right)\right)^2\hbar^4}.
    \end{aligned}
\end{equation}
The values of $a,b$, and $\delta$ are found by fitting $E_n$ with the experimental data. It is worth mentioning that the procedure presented can be generalized for any potential of Coulomb-like reciprocal terms plus a polynomial. Each added polynomial term will be accompanied by another expansion about $\left(\frac{1}{x}\right)^n$. Each added reciprocal term will just form into a polynomial after the substitution $x=\frac{1}{A}$. The same methods used to deduce \eqref{eq:5.20} can be followed and can then be solved using the Nikiforov-Uvarov method. It is also worth noting that for nonzero radial momentum, \eqref{eq:energyeigenvalues} gives complex energy eigenvalues. We will address this problem in later sections, but for now, we only look at the ${p_r}=0$ scenario.

\bigskip

\section{Mass Spectra of Heavy Quarks}
In this section, \eqref{eq:energyeigenvalues} will be compared to experimental results and other theoretical models for the mass spectrum of heavy quarks. The sign in \eqref{eq:energyeigenvalues} will be decided by the model having real values. Due to $E_n$ containing only three parameters, the model can be fit under three different conditions. Thus, for a generalized Cornell potential with additional terms, a more accurate model could possibly be generated. 
The mass spectra in three dimensions can be calculated via
\begin{equation}
    \begin{aligned}
        M=m_q+m_{\overline{q}}+E_{nl}
    \end{aligned}
\end{equation}
For calculation of the mass spectra, we look for the radial solutions of $E_{nl}$; thus, we let ${p_r}=0$. Using natural units, we also let $\hbar=1$.
\begin{equation}
\label{eq:6.2}
    \begin{aligned}
        M=m_q+m_{\overline{q}} + \frac{\begin{pmatrix}
            \frac{6bm^2\left(-2-4l\left(l+1\right)+n+n^2-\frac{4bm}{\delta^3}\right)^2}{\delta} \\
            -\frac{m\left(5+8l\left(l+1\right)+2n\left(n+1\right)+\frac{8bm}{\delta^3}\right)\left(-3bm+am\delta^2\right)^2}{\delta^4} \\
            \pm\sqrt{\frac{m^6\left(1+2n\right)^2\left(-3b+a\delta^2\right)^4(16bm+\left(9+16l+16l^2)\delta^3\right)\ }{\delta^{11}}}
        \end{pmatrix}}{2m^2\left(-2-4l\left(l+1\right)+n+n^2-\frac{4bm}{\delta^3}\right)^2}
    \end{aligned}
\end{equation}
where $m$ is the reduced mass and $m_q,m_{\overline{q}}$ are the masses of the quark and anti-quark respectively. The Cornell potential parameters are found by fitting \eqref{eq:6.2} to the experimental results of [8]. All results, including ones from other theoretical models ([9,10,11,12,13,14,15]), are recorded below. In our calculations, we use $m_c=1.23\ GeV,\ m_b=4.19 \ GeV$. The sign of $a$ refers to the charges of the two quarks. $b$ corresponds to the confinement force. $\delta$ is the value of the transformed position operator at the characteristic radius. The total error is found by taking the mean of the relative errors with respect to the experimental data [8] for each work.

\begin{table}[h]
\centering
\caption{Fitted Parameters for Various Mesons.\label{tab:1}}
\smallskip
\begin{tabular}{@{}llll@{}}
\toprule
Parameter & $c\overline{c}$&$b\overline{b}$&$b\overline{c}$\\
\midrule
$a$   & -2.5423 & -1.1820 & 116.66\\
$b$ $\left(GeV^2\right)$ & 0.4278 & 0.7912 & 0.5678\\
$\delta $ $\left(GeV\right)$ & 0.4286 & 0.6276 & 0.1778\\
\botrule
\end{tabular}
\end{table}

\begin{table}[h]
\centering
\caption{Mass Spectrum of $c\overline{c}$ Meson.\label{tab:2}}
\smallskip
\begin{tabular*}{\textwidth}{@{\extracolsep\fill}lccccccc}
\toprule
State & This Work & Exp. [8] & Ref. [9] & Ref. [10] & Ref. [11] & Ref. [12] & Ref. [13]\\
\midrule
1S & 3.097&  3.097& 3.098&  3.096 & 3.096&  3.096 & 3.097\\
2S &3.657&  3.686& 3.689 & 3.686&  3.686 & 3.686 & 3.773\\
1P &3.511 & 3.511 &3.262 & — & 3.527&  3.214&  3.511\\
2P &3.938 & 3.927& 3.784 & 3.757 & 3.687&  3.773&  3.927\\
3S &4.039 & 4.039 &4.041 & 4.323 & 4.040 & 4.275 & 4.039\\
4S &4.311 & — & 4.266 & 4.989 & 4.360 & 4.865 & 4.170\\
1D	&4.005 & 3.770 &3.515 & — & 3.098&  3.412 & 3.852\\
 Total Rel. Error & 1.22 \% & — & 2.94 \% & 2.84 \% & 4.07 \% & 4.75 \% & 0.76 \% \\
\botrule
\end{tabular*}
\end{table} 

\bigskip

\begin{table}[h]
\centering
\caption{Mass Spectrum of $b\overline{b}$ Meson.\label{tab:3}}
\smallskip
\begin{tabular*}{\textwidth}{@{\extracolsep\fill}lccccccc}
\toprule
State & This Work & Exp. [8] & Ref. [9] & Ref. [10] & Ref. [11] & Ref. [12] & Ref. [13] \\
\midrule
1S & 9.460 & 9.460&  9.46&  9.515 & 9.460 & 9.460  &9.460\\
2S & 9.975 & 10.023 & 10.023 & 10.018&  10.023&  10.023  & 10.114\\
1P & 9.746 & 9.899 & 9.608 & — & 9.661&  9.492  & 9.825\\
2P & 10.185&  10.260 & 10.110&  10.09&  10.238 & 10.038  & 10.260\\
3S & 10.355 & 10.355 & 10.365 & 10.441&  10.355 & 10.585  & 10.389\\
4S & 10.644 & 10.579 & 10.588 & 10.858&  10.567 & 11.148  & 10.530\\
1D & 10.164 & 10.164 & 9.841 & — & 9.943 & 9.551 &  10.164\\
Total Rel. Error & 0.47 \% & — & 1.11 \% & 1.15 \% & 0.70 \% & 2.84 \% & 0.34 \% \\
\botrule
\end{tabular*}
\end{table}

\begin{table}[h]
\centering
\caption{Mass Spectrum of $b\overline{c}$ Meson.\label{tab:4}}

\begin{tabular*}{\textwidth}{@{\extracolsep\fill}lccccccc}
\toprule
State & This Work & Exp. [8] & Ref. [9] & Ref. [14] & Ref. [15] & Ref. [12] & Ref. [13] \\
\midrule
1S  &6.275 & 6.275  &  6.274  &  6.277 & 6.268  &  6.277  &  6.275\\
2S  &6.842 & 6.842  &  6.845  &  6.496 & 6.895  &  6.814  &  6.842\\
1P  &6.360 & —  &  6.519  &  6.423  &6.529  &  6.340  &  6.336\\
2P  &6.919 & —  &  6.959  &  6.642  &7.156  &  6.851  &  6.889\\
3S  &7.356 & —  &  7.125  &  6.715  &7.522  &  7.351  &  7.282\\
4S & 7.822 & —  &  7.283  &  6.933  &—  &  7.889  &  7.631\\
1D	&  6.524 & —  &  6.813  &  6.569 & —  &  6.452  &  6.452\\
Total Rel. Error & 0.00 \% & — & 1.62 \% & 2.54 \% & 0.44 \% & 0.22 \% & 0.00 \% \\
\botrule
\end{tabular*}
\end{table}

\FloatBarrier

\bigskip
In Ref. [9], the mass spectra of the $c\overline{c},\ b\overline{b}$, and $b\overline{c}$ mesons were obtained for the combination of the inversely quadratic and Killingbeck potential via use of the WKB method for the SE. In Ref. [10], the mass spectra of the $c\overline{c},b\overline{b}$ mesons were computed using the SE with the SEM for the quark anti-quark interaction potential. In Ref. [11], the mass spectra of the $c\overline{c},b\overline{b}$ mesons were calculated using a class of solutions of the KGE for the Yukawa potential via the NU method. Ref. [12] used the AIM for the general action potential to obtain the mass spectra of $c\overline{c},b\overline{b}$, and $b\overline{c}$. Ref. [13] uses the eigenvalue solutions from the Dirac Equation for the generalized Cornell potential via NU method. In [15] the mass spectra of heavy mesons were obtained for the trigonometric Rosen-Morse potential. [14] finds the mass spectra using the EAIM and SEM for the extended Cornell potential.
\\

In case of the $c\overline{c}$ meson, our predictions for 1S, 1P, and 3S are exact to Exp. [8]. Our results for 2S and 2P differ with the experimental results by 29 MeV and 11 MeV respectively. The prediction of 1D is overestimated by 235 MeV, while the outcomes of Refs. [9],[11],[12],[13] differ with the experimental results by 255 MeV, 672 MeV, 358 MeV, and 82 MeV respectively. The value of 4S is within the reasonable bounds set by the results of Refs. [9-13] and is 219 MeV from the average of Refs. [9-13]. Taking into consideration the results of this paper, the only other model that is closer to the average of predictions for the 4S orbital is Ref. [11], which has a difference of 133 MeV, meanwhile this paper has a difference of 182 MeV. The sum of the differences between the predictions of the $c\overline{c}$ meson and Exp. [8] for this work and Refs. [9-13] is 275 MeV, 652 MeV, 455 MeV, 929 MeV, 1045 MeV, and 169 MeV respectively. It is also worth noting that Ref. [10] does not provide results for the 1P and 1D orbitals.
\\

In case of the $b\overline{b}$ meson, our predictions for 1S, 3S, and 1D are exact to Exp. [8]. Our results for 2S, 1P, 2P, and 4S differ to the experimental results by 48 MeV, 153 MeV, 75 MeV, and 65 MeV respectively. The sum of the differences between the predictions of the $b\overline{b}$ meson and Exp. [8] for this work and Refs. [9-13] is 341 MeV, 748 MeV, 595 MeV, 493 MeV, 2041 MeV, and 248 MeV respectively. It is once again worth noting that Ref. [10] doesn’t provide results for the 1P and 1D orbitals.  
\\

In case of the $b\overline{c}$ meson, our results for 1S and 2S are consistent with Exp. [8]. There is a lack of data to generate accuracy of the models with respect to experimental results. but the values of the model presented in this work are consistently in good accord with the results of Refs. [9],[12],[13],[14], and [15]. 
\\

Through this amalgam of results presented, suggests the validity of this work’s model presented and with that, the methods in Section 5. It is also worth mentioning that this work’s model consistently proves itself to more accurate than the models presented in Refs. [9-15] except for Ref. [13]. Ref. [13] uses an extended Cornell potential with three additional polynomial terms. This allows for more constraints upon curve fitting and more variability to model the data more accurately. Upon using a generalized or extended Cornell potential, the model in this work has a potential to be more accurate. \\

The graphs for the mass spectrum of each meson vs the principal quantum number, $n$, are given below:

\begin{figure}[H]
\centering
\includegraphics[width=.4\textwidth]{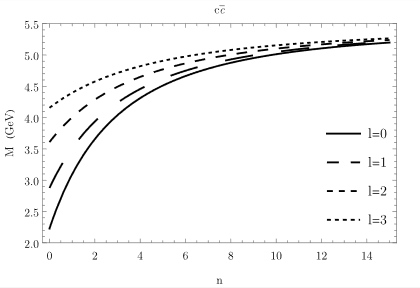}
\qquad
\includegraphics[width=.4\textwidth]{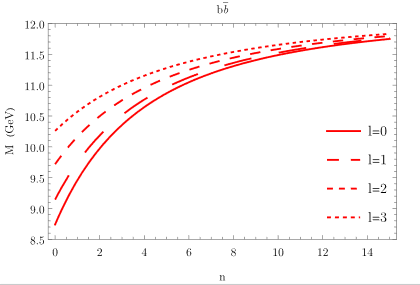}
\qquad
\includegraphics[width=.4\textwidth]{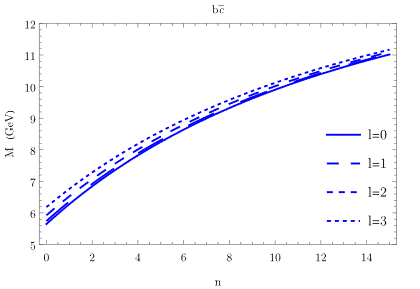}
\caption{meson mass spectrum vs. $n$, $c\overline{c}$ (right),  $b\overline{b}$ (right), $b\overline{c}$ (bottom) .\label{fig:1}}
\end{figure}

The graphs for the mass spectra of each meson vs. the orbital quantum number, $l$, are given below:

\begin{figure}[H]
\centering
\includegraphics[width=.4\textwidth]{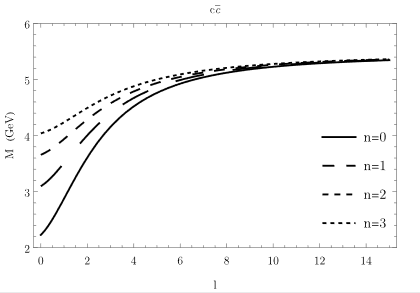}
\qquad
\includegraphics[width=.4\textwidth]{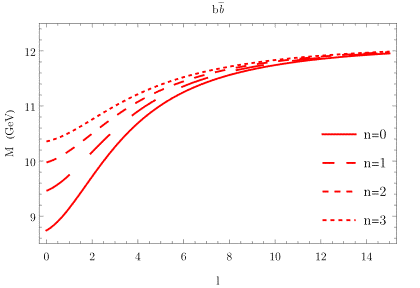}
\qquad
\includegraphics[width=.4\textwidth]{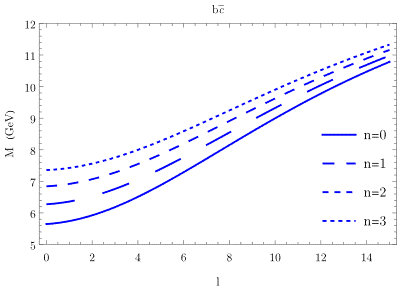}
\caption{meson mass spectrum vs. $n$, $c\overline{c}$ (right),  $b\overline{b}$ (right), $b\overline{c}$ (bottom) .\label{fig:2}}
\end{figure}

From these graphs, there is a consistent trend that an increase in $n$ and $l$ results in a greater mass spectrum. As each quantum number increases, the mass spectrum tapers off asymptotically to a specific value.
\\

To keep consistency with the charge of the quarks and anti-quarks, Cornell potential parameter $a$ is either chosen to be positive or negative. When fitting the data for the $b\overline{c}$ meson, $a$ is chosen to be positive. Comparing the magnitudes of parameter $a$ when it is positive vs. when it is negative, the positive parameter $a$ has a significantly larger magnitude. This trend is also seen when fitting the $c\overline{c}$ and $b\overline{b}$ mesons and suggests that same charged quarks have a higher QCD running coupling. This can be best shown by graphing the $b\overline{c}$ mass spectrum curve against the possible values that $a$ can attain. 

\begin{figure}[H]
\centering
\includegraphics[width=.5\textwidth]{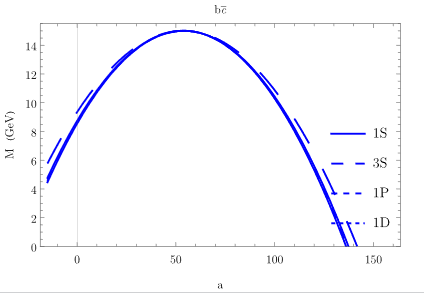}
\caption{$cc$ meson mass spectrum of $b\overline{c}$ vs. $a$.\label{fig:3}}
\end{figure}

It is shown in Figure \ref{fig:3} that for each energy level, there are two corresponding $a$ values. Because the graph is shifted to the right, any possible $a$ values that are negative will be much smaller than any a values that are positive. The graph is shifted far to the right for the $b\overline{c}$ meson compared to the $c\overline{c}$ and $b\overline{b}$ mesons, and since the general form of the mass spectrum doesn't change for the mesons, we can conclude that heavy quark anti-quark systems have a larger coupling constant $a$.
\smallskip
\section{Complex Energy Eigenvalues}

In this section, we analyze \eqref{eq:wavefunction} with the parameters found in section 6, and present new insight into calculating real energy eigenvalues and a real phase space wave function.

Looking at \eqref{eq:energyeigenvalues} with the parameter values found from the mass spectra, it is noticeable that for any radial momentum ${p_r}\neq0$, \eqref{eq:energyeigenvalues} produces complex energy eigenvalues. The complex energy eigenvalues suggest how the system does not persist in time with $e^{i\omega t}$ dependence, but decays exponentially. This phenomenon is accurate to the nature of a meson considering there is an energy threshold for the quark antiquark system to stay bound. Due to the theory of residues, an exponential decay in the p space is the result of an imaginary pole in the position space. For purely real position and momentum to be observed with complex eigenvalues, there must exist pairs of eigenvalues made of a complex number and its conjugate such that addition resolves along the real axis.

From this analysis, we speculate that a phase shift is being applied onto the wave function which results the eigenvalues not adding onto the real axis. When evaluating the ground state wave function by \eqref{eq:wavefunction}, we see that 
\begin{equation}
\label{eq:7.1}
    \begin{aligned}
        \psi\left(r,\overline{p}_r\right)=Be^\frac{\left(r+\frac{i\hbar\overline{p}_r}{2}\right)\left(2i{p_r}-\sqrt{-4{p_r}^2+\alpha\hbar^2}\right)}{\hbar}\left(\frac{1}{r+\frac{i\hbar\overline{p}_r}{2}}\right)^{-1+\frac{4i{p_r}+\beta\hbar}{2\sqrt{-4{p_r}^2+\alpha\hbar^2}}}
    \end{aligned}
\end{equation}
The inverse Laplace Transform of \eqref{eq:7.1} can be mapped to its Fourier counterpart through the mapping $s=i\omega$, where $s$ and $\omega$ are the Laplace and Fourier variables respectively. This inverse Fourier transform can be directly evaluated from $\overline{p}_r\rightarrow\ {p_r}$, yielding
\begin{equation}
    \begin{aligned}
        \psi\left(r,{p_r}\right)=\ F_{\left(\overline{p}_r\rightarrow\ {p_r}\right)}^{-1}\left[Be^\frac{\left(r+\frac{i\hbar\overline{p}_r}{2}\right)\left(2i{p_r}-\sqrt{-4{p_r}^2+\alpha\hbar^2}\right)}{\hbar}\left(\frac{1}{r+\frac{i\hbar\overline{p}_r}{2}}\right)^{-1+\frac{4i{p_r}+\beta\hbar}{2\sqrt{-4{p_r}^2+\alpha\hbar^2}}}\right].
    \end{aligned}
\end{equation}
\begin{equation}
\label{eq:7.3}
    \begin{aligned}
    &\psi\left(r,{p_r}\right)= \\
        &B\frac{\begin{pmatrix}
            5e^{r\left(2i{p_r}-\sqrt{-4{p_r}^2+\alpha}\right)+r\left(-2i{p_r}-2i{p_r}r+\sqrt{-4{p_r}^2+\alpha}\right)}\sqrt{\frac{2}{\pi}}\left(-\frac{1}{r}\right)^{-\frac{4i{p_r}+\beta}{2\sqrt{-4{p_r}^2+\alpha}}}r^{-\frac{4i{p_r}+\beta}{2\sqrt{-4{p_r}^2+\alpha}}} \\
            \begin{pmatrix}
                \left(2i{p_r}+2i{p_r}r-\sqrt{-4{p_r}^2+\alpha}\right)^\frac{4i{p_r}+\beta}{2\sqrt{-4{p_r}^2+\alpha}}-\left(-\frac{1}{r}\right)^\frac{4i{p_r}+\beta}{2\sqrt{-4{p_r}^2+\alpha}}r^\frac{4i{p_r}+\beta}{2\sqrt{-4{p_r}^2+\alpha}} \\
                \times\left(-2i{p_r}-2i{p_r}r+\sqrt{-4{p_r}^2+\alpha}\right)^\frac{4i{p_r}+\beta}{2\sqrt{-4{p_r}^2+\alpha}}
            \end{pmatrix} \\
            \times{}\left(-2i\pi\sqrt{-4{p_r}^2+\alpha}\ \ Csch\left(\frac{\pi\left(4{p_r}-i\beta\right)}{2\sqrt{-4{p_r}^2+\alpha}}\right)-\begin{pmatrix}
                \left(-4i{p_r}+4\sqrt{-4{p_r}^2+\alpha}-\beta\right) \\
                \times\begin{pmatrix}
                    \Gamma\left(\frac{4i{p_r}-4\sqrt{-4{p_r}^2+\alpha}+\beta}{2\sqrt{-4{p_r}^2+\alpha}}\right) \\
                -\Gamma\begin{pmatrix}
                    2-\frac{2i{p_r}}{\sqrt{-4{p_r}^2+\alpha}}-\frac{\beta}{2\sqrt{-4{p_r}^2+\alpha}},r-2i{p_r} \\
                    -2i{p_r}r+\sqrt{-4{p_r}^2+\alpha}
                \end{pmatrix}\end{pmatrix}
            \end{pmatrix}\right)
        \end{pmatrix}}{\sqrt{-4{p_r}^2+\alpha}\ \left(-2i{p_r}-2i{p_r}r+\sqrt{-4{p_r}^2+\alpha}\right)^2\Gamma\left(-1+\frac{4i{p_r}+\beta}{2\sqrt{-4{p_r}^2+\alpha}}\right)}
    \end{aligned}
\end{equation}

From \eqref{eq:7.3} we can see a factor of $e^{-2i{p_r}r}$, which would be $e^\frac{-2i{p_r}r}{\hbar}$ without $\hbar=1$. This same factor is also found within all evaluated wave functions past $n=0$. Thus, using this knowledge and our previous assumption about the complex energy eigenvalues, we make the ansatz
\begin{equation}
    \begin{aligned}
        \psi\left(r,{p_r}\right)=e^{-\frac{2i{p_r}r}{\hbar}}\ \Omega\left(r,{p_r}\right)
    \end{aligned}
\end{equation}
The phase space Schrodinger equation then becomes
\begin{equation}
\label{eq:7.5}
    \begin{aligned}
        \begin{pmatrix}
            
        \frac{l\left(1+l\right)\hbar^2 \Omega\left(r,\overline{p}_r\right)}{2m\left(\frac{i\hbar}{2}\overline{p}_r+r\right)^2}+\frac{a\ \Omega\left(r,\overline{p}_r\right)}{r+\frac{i\hbar\overline{p}_r}{2}}+b\left(r+\frac{i\hbar\overline{p}_r}{2}\right)\Omega\left(r,\overline{p}_r\right) \\
        -\frac{\hbar^2}{2m\left(r+\frac{i\hbar\overline{p}_r}{2}\right)}\frac{\partial\Omega\left(r,\overline{p}_r\right)}{\partial r}-\frac{\hbar^2}{8m}\frac{\partial^2\Omega\left(r,\overline{p}_r\right)}{\partial r^2}\end{pmatrix}=E_n\Omega\left(r,\overline{p}_r\right).
    \end{aligned}
\end{equation}
The origin of this rotational factor is unknown, and its nature will be further investigated in upcoming papers. Once again, the substitution $A=r+\frac{i\hbar\overline{p}_r}{2}$ is made where the chain rule yields
\begin{equation}
    \frac{\partial\Omega\left(r,\overline{p}_r\right)}{\partial r}=\frac{\partial\Omega(A)}{\partial\ A}\frac{\partial\ A}{\partial r}=\frac{\partial\Omega\left(A\right)}{\partial r},\ \ \frac{\partial^2\Omega\left(r,\overline{p}_r\right)}{\partial r^2}=\frac{\partial^2\Omega(A)}{\partial\ A^2}\frac{\partial\ A^2}{\partial r^2}=\frac{\partial^2\Omega\left(A\right)}{\partial r^2}.
\end{equation}
\eqref{eq:7.5} then becomes
\begin{equation}
    \begin{aligned}
        \left(\frac{a}{A}+Ab-E_n+\frac{l\left(l+1\right)}{2A^2m}\right)\Omega\left(A\right)-\frac{\hbar^2\Omega^\prime\left(A\right)}{2mA}-\frac{\hbar^2\Omega^{\prime\prime}(A)}{8m}=0.
    \end{aligned}
\end{equation}
Making the substitution $x=\frac{1}{A}$, the chain rule yields 
\begin{equation}
    \begin{aligned}
        \frac{d\Omega(A)}{dA}=\frac{d\Omega(x)}{dx}\frac{dx}{dA}=\ -\frac{1}{A^2}\frac{d\Omega\left(x\right)}{dx},\ \ \frac{d^2\Omega\left(A\right)}{dA^2}=\frac{d^2\Omega\left(x\right)}{dx^2}\frac{dx^2}{dA^2}=\frac{1}{A^4}\frac{d^2\Omega\left(x\right)}{dx^2}.
    \end{aligned}
\end{equation}
Now,
\begin{equation}
\label{eq:7.9}
    \begin{aligned}
        \left(ax+\frac{b}{x}-E_n+\frac{l\left(l+1\right)x^2}{2m}\right)\Omega\left(x\right)+\frac{\hbar^2\Omega^\prime\left(x\right)x^3}{2m}-\frac{x^4\hbar^2\Omega^{\prime\prime}(x)}{8m}=0
    \end{aligned}
\end{equation}
We employ the same approximation scheme about $x=\frac{1}{A}$. We let $y=x-\delta $ around the singularity $y=0$ and generate the power series to second degree: 
\begin{equation}
\label{eq:7.10}
    \begin{aligned}
        \left(\frac{1}{x}\right)=\frac{1}{\left(y+\delta\right)}=\left(\frac{1}{\delta}\right)\left(\frac{1}{1+\frac{y}{\delta}}\right)=\ \frac{3}{\delta}-\frac{3x}{\delta^2}+\frac{x^2}{\delta^3}
    \end{aligned}
\end{equation}
Substituting \eqref{eq:7.10} into \eqref{eq:7.9}, 
\begin{equation}
\label{eq:7.11}
    \begin{aligned}
        \left(ax+b\left(\frac{3}{\delta}-\frac{3x}{\delta^2}+\frac{x^2}{\delta^3}\right)-E_n+\frac{l\left(l+1\right)x^2}{2m}\right)\Omega\left(x\right) + \frac{\hbar^2\Omega^\prime\left(x\right)x^3}{2m}-\frac{x^4\hbar^2\Omega^{\prime\prime}(x)}{8m}=0
    \end{aligned}
\end{equation}
Creating the definition for constants $\alpha, \beta, \gamma$,
\begin{equation}
    \begin{aligned}
        \alpha\equiv\frac{8E_nm}{\hbar^2}-\frac{24bm}{\delta\hbar^2},\ \ \beta\equiv\ -\frac{8am}{\hbar^2}+\frac{24bm}{\delta^2\hbar^2},\ \ \gamma\equiv-4L-4L^2-\frac{8bm}{\hbar^2\delta^3},
    \end{aligned}
\end{equation}
\eqref{eq:7.11} can be written as
\begin{equation}
    \begin{aligned}
        \frac{\left(\alpha+\beta x+\gamma x^2\right)\Omega\left(x\right)}{x^4}+\frac{-4x{\Omega}^\prime\left(x\right)}{x^2}+\Omega^{\prime\prime}(x)=0.
    \end{aligned}
\end{equation}
We can now apply the Nikiforov-Uvarov method once again:
\begin{equation}
    \begin{aligned}
        \sigma\left(x\right)=x^2, \quad \widetilde{\sigma}\left(x\right)=\alpha+x\beta+x^2\gamma, \quad \widetilde{\tau}\left(x\right)=-4x
    \end{aligned}
\end{equation}
From \eqref{eq:4.7},
\begin{equation}
    \begin{aligned}
        \pi\left(x\right)=3x\pm\sqrt{9x^2+Kx^2-\alpha-x\beta-s^2\gamma}
    \end{aligned}
\end{equation}
Since the polynomial under the radical must a square of a polynomial, the discriminant must equal zero.
\begin{equation}
    \begin{aligned}
        \left(-\beta\right)^2-4\left(-\alpha\right)\left(9+K-\gamma\right)=0
    \end{aligned}
\end{equation}
\begin{equation}
    \begin{aligned}
        K=\frac{-36\alpha-\beta^2+4\alpha\gamma}{4\alpha}
    \end{aligned}
\end{equation}
Thus,
\begin{equation}
\label{eq:7.18}
    \begin{aligned}
        \pi\left(x\right)=3x\pm\frac{2\alpha+x\beta}{2\sqrt{-\alpha}}.
    \end{aligned}
\end{equation}
For the rest of this paper, the choice of using the plus sign in \eqref{eq:7.18} will be used in defining $\pi(x)$ in order to ensure its derivative is $\geq{0}$. From \eqref{eq:4.4}, $\phi(x)$ can be found by the differential equation
\begin{equation}
    \begin{aligned}
        \frac{\phi^\prime\left(x\right)}{\phi(x)}=\frac{\pi\left(x\right)}{\sigma(x)}=\frac{3x-\frac{2\alpha+x\beta}{2\sqrt{-\alpha}}}{x^2},
    \end{aligned}
\end{equation}
which yields the exponential solution
\begin{equation}
    \begin{aligned}
        \phi\left(x\right)=e^{-\frac{\sqrt{-\alpha}}{x}}\ x^{3-\frac{\beta}{2\sqrt{-\alpha}}}.
    \end{aligned}
\end{equation}
From \eqref{eq:4.10},
\begin{equation}
    \begin{aligned}
        \tau\left(x\right)=\widetilde{\tau}\left(x\right)+2\pi\left(x\right)=2\sqrt{-\alpha}+x\left(2+\frac{\alpha\beta}{\left(-\alpha\right)^\frac{3}{2}}\right)
    \end{aligned}
\end{equation}
\eqref{eq:4.6} is used to find the weight function $\rho\left(x\right)$, where 
\begin{equation}
    \begin{aligned}
        \left[\sigma\left(x\right)\rho\left(x\right)\right]^\prime=\tau\left(x\right)\rho\left(x\right)
    \end{aligned}
\end{equation}
\begin{equation}
    \begin{aligned}
        2x\rho\left(x\right)+x^2\rho^\prime\left(x\right)=\left(2\sqrt{-\alpha}+x\left(2+\frac{\alpha\beta}{\left(-\alpha\right)^\frac{3}{2}}\right)\right)\rho\left(x\right)
    \end{aligned}
\end{equation}
\begin{equation}
    \begin{aligned}
        \rho\left(x\right)=e^\frac{\frac{2\alpha}{x}-\left(\beta+2\sqrt{-\alpha}\left(-1+\hbar\right)\right)\ln{x}}{\sqrt{-\alpha}\hbar}
    \end{aligned}
\end{equation}
With $\rho(x)$ and $\sigma\left(x\right)$, the function $y$ can be formulated as
\begin{equation}
    \begin{aligned}
        y_n\left(x\right)=\frac{B}{\rho\left(x\right)}\frac{d^n}{dx^n}\left(\sigma^n\left(x\right)\rho\left(x\right)\right)
    \end{aligned}
\end{equation}
Where $B$ is a normalization constant. Recalling \eqref{eq:4.2}, the wave function must then be 
\begin{equation}
\label{eq:7.26}
    \begin{aligned}
        \psi_n\left(x\right)=e^{-\frac{2i{p_r}r}{\hbar}}\ \Omega_n(x)={e^{-\frac{2i{p_r}r}{\hbar}}y}_n\left(x\right)\phi\left(x\right)=e^{-\frac{2i{p_r}r}{\hbar}}\phi\left(x\right)\frac{B}{\rho\left(x\right)}\ \frac{d^n}{dx^n}\left(\sigma^n\left(x\right)\rho\left(x\right)\right)
    \end{aligned}
\end{equation}
Back-substituting $x\rightarrow\frac{1}{A}=\frac{1}{r+\frac{i\hbar}{2}\overline{p}_r}$ , the term $\phi\left(x\right)\frac{B}{\rho\left(x\right)}$ becomes
\begin{equation}
\label{eq:7.27}
    \begin{aligned}
        \phi(x)\frac{B}{\rho\left(x\right)}&=\ Be^{-\frac{\sqrt{-\alpha}\left(-2+\hbar\right)}{x\hbar}}x^{5-\frac{2}{\hbar}-\frac{\beta\left(-2+\hbar\right)}{2\sqrt{-\alpha}\hbar}} \\
        &=Be^{-\frac{\sqrt{-\alpha}\left(-2+\hbar\right)}{\left(\frac{1}{r+\frac{i\hbar}{2}\overline{p}_r}\right)\hbar}}\left(\frac{1}{r+\frac{i\hbar}{2}\overline{p}_r}\right)^{5-\frac{2}{\hbar}-\frac{\beta\left(-2+\hbar\right)}{2\sqrt{-\alpha}\hbar}}.
    \end{aligned}
\end{equation}
The term $\frac{d^n}{dx^n}\left(\sigma^n\left(x\right)\rho\left(x\right)\right)$ can also undergo the substitution $x\rightarrow\frac{1}{A}=\frac{1}{r+\frac{i\hbar}{2}\overline{p}_r}$ to yield
\begin{equation}
\label{eq:7.28}
    \begin{aligned}
        \frac{d^n}{dx^n}\left(\sigma^n\left(x\right)\rho\left(x\right)\right)&=\frac{d^n}{d\left(\frac{1}{r+\frac{i\hbar\overline{p}_r}{2}}\right)^n}\left(\sigma^n\left(\frac{1}{r+\frac{i\hbar\overline{p}_r}{2}}\right)\rho\left(\frac{1}{r+\frac{i\hbar\overline{p}_r}{2}}\right)\right) \\
        &=\frac{d^n}{d\left(\frac{1}{r+\frac{i\hbar\overline{p}_r}{2}}\right)^n}\left(\left(\frac{1}{r+\frac{i\hbar\overline{p}_r}{2}}\right)^2e^\frac{\frac{2\alpha}{\frac{1}{r+\frac{i\hbar\overline{p}_r}{2}}}-\left(\beta+2\sqrt{-\alpha}\left(-1+\hbar\right)\right)\ln{\left(\frac{1}{r+\frac{i\hbar\overline{p}_r}{2}}\right)}}{\sqrt{-\alpha}\hbar}\right)
    \end{aligned}
\end{equation}
Substituting \eqref{eq:7.28} and \eqref{eq:7.27} back into \eqref{eq:7.26}:
\begin{equation}
    \begin{aligned}
        \psi_n\left(r,\overline{p}_r\right)&= \\
        &e^{-\frac{2i{p_r}r}{\hbar}}\begin{pmatrix}
            Be^{-\frac{\sqrt{-\alpha}\left(-2+\hbar\right)}{\left(\frac{1}{r+\frac{i\hbar}{2}\overline{p}_r}\right)\hbar}}\left(\frac{1}{r+\frac{i\hbar}{2}\overline{p}_r}\right)^{5-\frac{2}{\hbar}-\frac{\beta\left(-2+\hbar\right)}{2\sqrt{-\alpha}\hbar}} \\
            \times\frac{d^n}{d\left(\frac{1}{r+\frac{i\hbar\overline{p}_r}{2}}\right)^n}\left(\left(\frac{1}{r+\frac{i\hbar\overline{p}_r}{2}}\right)^2e^\frac{\frac{2\alpha}{\frac{1}{r+\frac{i\hbar\overline{p}_r}{2}}}-\left(\beta+2\sqrt{-\alpha}\left(-1+\hbar\right)\right)\ln{\left(\frac{1}{r+\frac{i\hbar\overline{p}_r}{2}}\right)}}{\sqrt{-\alpha}\hbar}\right)
        \end{pmatrix}.
    \end{aligned}
\end{equation}
Now, the wave function in $r$ and ${p_r}$ is:
\begin{equation}
\label{eq:wavefunction 2}
    \begin{aligned}
        \psi_n\left(r,{p_r}\right)&=
        &L_{\left(\overline{p}_r\rightarrow\ {p_r}\right)}^{-1}\begin{bmatrix}
            e^{-\frac{2i{p_r}r}{\hbar}}\begin{pmatrix}
            Be^{-\frac{\sqrt{-\alpha}\left(-2+\hbar\right)}{\left(\frac{1}{r+\frac{i\hbar}{2}\overline{p}_r}\right)\hbar}}\left(\frac{1}{r+\frac{i\hbar}{2}\overline{p}_r}\right)^{5-\frac{2}{\hbar}-\frac{\beta\left(-2+\hbar\right)}{2\sqrt{-\alpha}\hbar}} \\
            \times\frac{d^n}{d\left(\frac{1}{r+\frac{i\hbar\overline{p}_r}{2}}\right)^n}\left(\left(\frac{1}{r+\frac{i\hbar\overline{p}_r}{2}}\right)^2e^\frac{\frac{2\alpha}{\frac{1}{r+\frac{i\hbar\overline{p}_r}{2}}}-\left(\beta+2\sqrt{-\alpha}\left(-1+\hbar\right)\right)\ln{\left(\frac{1}{r+\frac{i\hbar\overline{p}_r}{2}}\right)}}{\sqrt{-\alpha}\hbar}\right)
        \end{pmatrix}
        \end{bmatrix}.
    \end{aligned}
\end{equation}
The corresponding energy eigenvalues are found via \eqref{eq:4.8} and \eqref{eq:4.9}.
\begin{equation}
    \begin{aligned}
        \lambda=K+\pi^\prime\left(x\right)=\ 3-\frac{\beta}{2\sqrt{-\alpha}}+\frac{-36\alpha-\beta^2+4\alpha\gamma}{4\alpha}
    \end{aligned}
\end{equation}
\begin{equation}
    \begin{aligned}
        \lambda_n= -n\tau^\prime\left(x\right)-\frac{n\left(n-1\right)\sigma^{\prime\prime}\left(x\right)}{2}=-n\left(n-1\right)+n\left(2+\frac{\alpha\beta}{\left(-\alpha\right)^\frac{3}{2}}\right)
    \end{aligned}
\end{equation}
Using the condition $\lambda=\lambda_n$,
\begin{equation}
\label{eq:7.33}
    \begin{aligned}
        3-\frac{\beta}{2\sqrt{-\alpha}}+\frac{-36\alpha-\beta^2+4\alpha\gamma}{4\alpha}\ =\ -n\left(n-1\right)+n\left(2+\frac{\alpha\beta}{\left(-\alpha\right)^\frac{3}{2}}\right)
    \end{aligned}
\end{equation}
\bigskip
$\alpha\equiv\frac{8E_nm}{\hbar^2}-\frac{24bm}{\delta\hbar^2}$. Substituting this into \eqref{eq:7.33} and solving for $E_n$ yields
\begin{equation}
    \begin{aligned}
        E_n=\frac{3b}{\delta}+\frac{-\beta^2\left(13+2n\left(n+1\right)-2\gamma\right)\hbar^2+\sqrt{\left(1-2n\right)^2\beta^4\left(25+8n-4\gamma\right)\hbar^4}}{64m\left(-6+\left(n-3\right)n+\gamma\right)^2}.
    \end{aligned}
\end{equation}
Substituting the values $\beta=\ -\frac{8am}{\hbar^2}+\frac{24bm}{\delta^2\hbar^2},\ \gamma=-4L-4L^2-\frac{8bm}{\hbar^2\delta^3}$, where m is the reduced mass, generates 
\begin{equation}
    \begin{aligned}
        E_n=\frac{3b}{\delta}+\frac{\begin{pmatrix}
            \sqrt{\frac{m^4\left(1-2n\right)^2\left(-3b+a\delta^2\right)^4\left(32bm+\left(25+16L\left(L+1\right)+8n\right)\delta^3\hbar^2\right)}{\delta^{11}\hbar^6}} \\
            \frac{m^2\left(-3b+a\delta^2\right)^2\left(-16bm-\left(13+8L\left(1+L\right)+2n\left(1+n\right)\right)\delta^3\hbar^2\right)}{\delta^7\hbar^4}
        \end{pmatrix}}{m\left(6+4L\left(1+L\right)-\left(n-3\right)n+\frac{8bm}{\delta^3\hbar^2}\right)}.
    \end{aligned}
\end{equation}
Thus, the wave function and energy eigenvalues can now be represented purely on the real plane. It is worth noting that with this change, the mass spectrum calculations become significantly more accurate. 
\\

To prevent confusion on how we curve fit the data with $p_r \neq 0$, we emphasize that $\delta$ is not dependent on $p_r$. $\delta$ is a constant that is dependent on a characteristic value (value obtained momentarily before particle collision) for $r$, and a characteristic value of $\overline{p}_r$. In other words, we assume that there is a particular value of position that appears when the particle and anti-particle obtain momentarily before collision. We also assume there is a particular value that the transformed momentum, $\overline{p}_r$, takes before collision. Combining these assumptions, we are stating that there is a constant value obtained by the transformed position operator before the collision of the particle and anti-particle. This value of A, the transformed position operator, is equal to
$\frac{1}{\delta}$, in which $\delta = \frac{1}{r + \frac{i\hbar}{2}\overline{p}_r}$.
\\

Compared to the other theoretical models previously discussed and the previous equation in this work, the new mass spectrum results is much more accurate and is on par with [13] which uses the Dirac equation over the Schrodinger Equation. The same quantum number trends are also seen. The new mass spectrum data is given below.
\begin{table}[h]
\centering
\caption{New Fitted Parameters for Various Mesons.\label{tab:5}}
\smallskip
\begin{tabular}{@{}llll@{}}
\toprule
Parameter&$c\overline{c}$&$b\overline{b}$&$b\overline{c}$\\
\midrule
$a$  & -1.6808 & -0.7383 & 105.67\\
$b$ $\left(GeV^2\right)$ & 0.4069 & 1.0628 & 0.5157\\
$\delta $ $\left(GeV\right)$ & 0.5074 & 1.1871 & 0.1763\\
\botrule
\end{tabular}
\end{table}

\begin{table}[h]
\centering
\caption{Mass Spectrum of $c\overline{c}$ Meson.\label{tab:6}}
\smallskip
\begin{tabular*}{\textwidth}{@{\extracolsep\fill}lccccccc}

\toprule
State & This Work & Exp. [8] & Ref. [9] & Ref. [10] & Ref. [11] & Ref. [12] & Ref. [13] \\
\midrule
1S & 3.097&  3.097& 3.098&  3.096 & 3.096&  3.096 & 3.097\\
2S &3.686&  3.686& 3.689 & 3.686&  3.686 & 3.686 & 3.773\\
1P &3.511 & 3.511 &3.262 & — & 3.527&  3.214&  3.511\\
2P &3.912 & 3.927& 3.784 & 3.757 & 3.687&  3.773&  3.927\\
3S &4.022 & 4.039 &4.041 & 4.323 & 4.040 & 4.275 & 4.039\\
4S &4.231 & — & 4.266 & 4.989 & 4.360 & 4.865 & 4.170\\
1D	&3.939 & 3.770 &3.515 & — & 3.098&  3.412 & 3.852\\
Total Rel. Error & 0.88 \% & — & 2.94 \% & 2.84 \% & 4.07 \% & 4.75 \% & 0.76 \% \\
\botrule
\end{tabular*}
\end{table}

\begin{table}[h]
\centering
\caption{Mass Spectrum of $b\overline{b}$ Meson.\label{tab:7}}
\smallskip
\begin{tabular*}{\textwidth}{@{\extracolsep\fill}lccccccc}

\toprule
State & This Work & Exp. [8] & Ref. [9] & Ref. [10] & Ref. [11] & Ref. [12] & Ref. [13] \\
\midrule
1S & 9.460 & 9.460&  9.46&1  9.515 & 9.460 & 9.460  &9.460\\
2S & 10.042 & 10.023 & 10.023 & 10.018&  10.023&  10.023  & 10.114\\
1P & 9.899 & 9.899 & 9.608 & — & 9.661&  9.492  & 9.825\\
2P & 10.268&  10.260 & 10.110&  10.09&  10.238 & 10.038  & 10.260\\
3S & 10.355 & 10.355 & 10.365 & 10.441&  10.355 & 10.585  & 10.389\\
4S & 10.542 & 10.579 & 10.588 & 10.858&  10.567 & 11.148  & 10.530\\
1D & 10.307 & 10.164 & 9.841 & — & 9.943 & 9.551 &  10.164\\
Total Rel. Error & 0.29 \% & —& 1.11 \% & 1.15 \% & 0.70 \% & 2.84 \% & 0.34 \% \\
\botrule
\end{tabular*}
\end{table}

\begin{table}[h]
\centering
\caption{Mass Spectrum of $b\overline{c}$ Meson.\label{tab:8}}
\smallskip
\begin{tabular*}{\textwidth}{@{\extracolsep\fill}lccccccc}
\toprule
State & This Work & Exp. [8] & Ref. [9] & Ref. [14] & Ref. [15] & Ref. [12] & Ref. [13] \\
\midrule
1S  &6.275 & 6.275  &  6.274  &  6.277 & 6.268  &  6.277  &  6.275\\
2S  &6.842 & 6.842  &  6.845  &  6.496 & 6.895  &  6.814  &  6.842\\
1P  &6.360 & —  &  6.519  &  6.423  &6.529  &  6.340  &  6.336\\
2P  &6.919 & —  &  6.959  &  6.642  &7.156  &  6.851  &  6.889\\
3S  &7.356 & —  &  7.125  &  6.715  &7.522  &  7.351  &  7.282\\
4S & 7.822 & —  &  7.283  &  6.933  &—  &  7.889  &  7.631\\
1D	&  6.524 & —  &  6.813  &  6.569 & —  &  6.452  &  6.452\\
Total Rel. Error & 0.00 \% & — & 1.62 \% & 2.54 \% & 0.44 \% & 0.22 \% & 0.00 \% \\
\botrule
\end{tabular*}
\end{table} 

\FloatBarrier

\section{Wave Function Evaluation}
In this section, the evaluation of wave functions given by \eqref{eq:wavefunction 2} will be shown. We start with the ground state wave function $n=0$, where
\begin{equation}
\label{eq:8.1}
    \begin{aligned}
        \psi\left(r,{p_r}\right) =L_{\left(\overline{p}_r\rightarrow {p_r}\right)}^{-1}\left[e^{-\frac{2i{p_r}r}{\hbar}}\left(Be^{-\sqrt{-\alpha}\left(r+\frac{i\overline{p}_r\hbar}{2}\right)}\left(\frac{1}{r+\frac{i\overline{p}_r\hbar}{2}}\right)^{3-\frac{\beta}{2\sqrt{-\alpha}}}\right)\right].
    \end{aligned}
\end{equation}
It is worth noting that when we state that $n=0$ is the ground state wave function, $n$ is the polynomial index from the Nikiforov-Uvarov method, not the quantum principle number. Using the transformation variable relationship $s=i\omega$, \eqref{eq:8.1} is mapped to the Fourier transform
\begin{equation}
    \begin{aligned}
        \psi\left(r,{p_r}\right) &=F_{\left(\overline{p}_r\rightarrow {p_r}\right)}^{-1}\left[e^{-\frac{2i{p_r}r}{\hbar}}\left(Be^{-\sqrt{-\alpha}\left(r-\frac{\overline{p}_r\hbar}{2}\right)}\left(\frac{1}{r-\frac{\overline{p}_r\hbar}{2}}\right)^{3-\frac{\beta}{2\sqrt{-\alpha}}}\right)\right]
        \\
        &=-e^{-2i{p_r}r+r\left(-2i{p_r}+\sqrt{-\alpha}\right)-r\sqrt{-\alpha}}\sqrt{\frac{2}{\pi}}\begin{pmatrix}
            \begin{pmatrix}
                \left(-\frac{1}{r}\right)^\frac{\beta}{2\sqrt{-\alpha}}r^\frac{\beta}{2\sqrt{-\alpha}}\left(2i{p_r}-\sqrt{-\alpha}\right)^{-\frac{\beta}{2\sqrt\alpha}} \\
                -\left(-2i{p_r}+\sqrt{-\alpha}\right)^{-\frac{\beta}{2\sqrt{-\alpha}}}
            \end{pmatrix}
            \\
            \times\left(-2i{p_r}+\sqrt{-\alpha}\right)^2\Gamma\left(-2+\frac{\beta}{2\sqrt{-\alpha}},r\left(-2i{p_r}+\sqrt{-\alpha}\right)\right)
        \end{pmatrix}.
    \end{aligned}
\end{equation}
Where the constants $\alpha,\beta,\gamma$ are given by $\alpha\equiv\frac{8E_nm}{\hbar^2}-\frac{24bm}{\delta\hbar^2}, \ \beta\equiv\ -\frac{8am}{\hbar^2}+\frac{24bm}{\delta^2\hbar^2},\ \gamma\equiv-4L-4L^2-\frac{8bm}{\hbar^2\delta^3}$. The normalization constant, $B$, is calculated via the condition
\begin{equation}
\label{eq:8.3}
    \begin{aligned}
        \int_{-\infty}^{\infty}\int_{0}^{\pi}\int_{0}^{2\pi}\int_{0}^{\infty}{\left|\psi\left(r,{p_r}\right)\right|^2\ r^2sin\theta\ drd\theta\ d\phi\ d{p_r}=1}.
    \end{aligned}
\end{equation}

Once again, as in the 1D scenario, \eqref{eq:8.3} does not include the angular momentum variables for $\phi$ and $\theta$ since the Schrodinger equation was angularly separated. Thus, ${p_r}$ only represents the radial momentum. The value for $B$ given by \eqref{eq:8.3} can by analytically or numerically determined.
\\

For the ground state wave function $n=0$ with 1S energy level, $B$ converges to 26, resulting in the normalized $c\overline{c}$ meson wave function amplitude and probability density:

\begin{figure}[H]
\centering
\includegraphics[width=.4\textwidth]{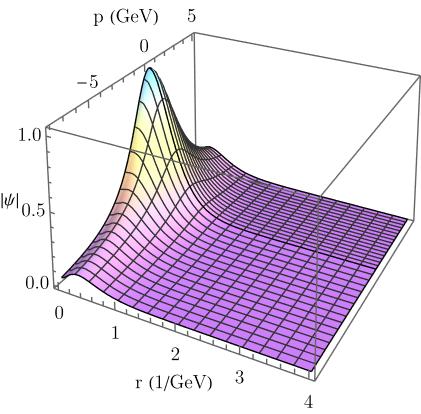}
\qquad
\includegraphics[width=.4\textwidth]{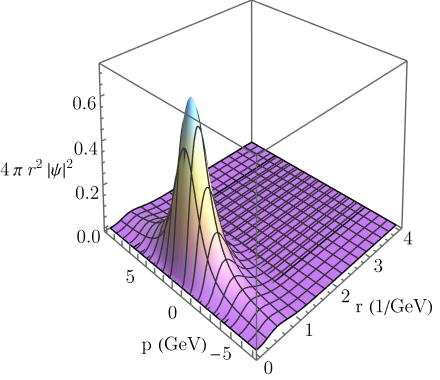}
\caption{1S $cc$ meson wave function probability amplitude(left) and density(right).\label{fig:4}}
\end{figure}

For the ground state wave function $n=0$ with 1P energy level, the normalization constant $B$ converges to 2411, and \eqref{eq:wavefunction 2} yields the $c\overline{c}$ meson wave function amplitude and probability density:

\begin{figure}[H]
\centering
\includegraphics[width=.4\textwidth]{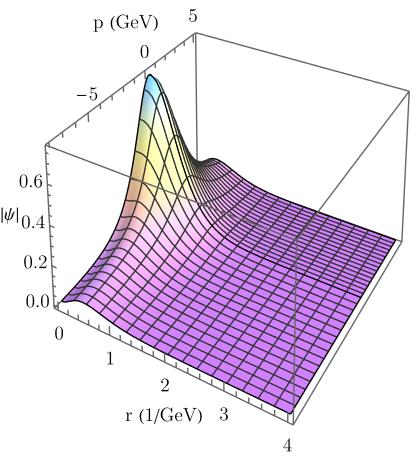}
\qquad
\includegraphics[width=.4\textwidth]{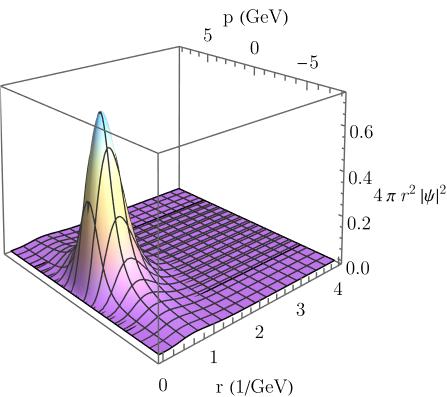}
\caption{1P $cc$ meson wave function probability amplitude(left) and density(right).\label{fig:5}}
\end{figure}

For the excited $n=1$ meson wave function, 
\begin{equation}
\label{eq:8.4}
    \begin{aligned}
        \psi\left(r,{p_r}\right) =L_{\left(\overline{p}_r\rightarrow\ {p_r}\right)}^{-1}\left[\frac{e^{-\frac{2i{p_r}r}{\hbar}}\left(e^{-\left(\left(\frac{i\overline{p}_r}{2}+r\right)\sqrt{-\alpha}\right)}\right)\left(\frac{1}{\frac{i\overline{p}_r}{2}+r}\right)^{3-\frac{\beta}{2\sqrt{-\alpha}}}\left(\frac{2\sqrt{-\alpha}}{\frac{i\overline{p}_r}{2}+r}-2\alpha-\frac{\beta}{\frac{i\overline{p}_r}{2}+r}\right)}{\sqrt{-\alpha}}\right].
    \end{aligned}
\end{equation}
Mapping \eqref{eq:8.4} to its Fourier counterpart, it can be directly evaluated to yield
\begin{equation}
    \begin{aligned}
        &\psi\left(r,{p_r}\right)= \\
        &2e^{-4\left(4i{p_r}+\sqrt{-\alpha}\right)}\begin{pmatrix}
            \left(\left(-\frac{1}{r}\right)^\frac{\beta}{2\sqrt{-\alpha}}r^\frac{\beta}{2\sqrt{-\alpha}}\left(2i{p_r}-\sqrt{-\alpha}\right)^{-\frac{\beta}{2\sqrt{-\alpha}}}-\left(2i{p_r}+\sqrt{-\alpha}\right)^{-\frac{\beta}{2\sqrt{-\alpha}}}\right) \\
            \times\left(2i{p_r}+\sqrt{-\alpha}\right)^2 \begin{pmatrix}
                -\frac{1}{\pi}2e^{r\left(-2i{p_r}+\sqrt{-\alpha}\right)} \\
                \times\left(\left(-\frac{1}{r}\right)^\frac{\beta}{2\sqrt{-\alpha}}r^\frac{\beta}{2\sqrt{-\alpha}}\left(2i{p_r}-\sqrt{-\alpha}\right)^{-\frac{\beta}{2\sqrt{-\alpha}}}-\left(2i{p_r}+\sqrt{-\alpha}\right)^{-\frac{\beta}{2\sqrt\alpha}}\right)
                \\
                \times\left(-2i{p_r}+\sqrt{-\alpha}\right)^4\beta\ \Gamma^2\left(-3+\frac{\beta}{2\sqrt{-\alpha}},r\left(-2i{p_r}+\sqrt{-\alpha}\right)\right)
                \\
                +{\ e}^{r\sqrt{-\alpha}}\sqrt{\frac{2}{\pi}}\alpha\frac{\Gamma\left(-2+\frac{\beta}{2\sqrt{-\alpha}},r\left(-2i{p_r}+\sqrt{-\alpha}\right)\right)}{\sqrt{-\alpha}}
            \end{pmatrix}
        \end{pmatrix},
    \end{aligned}
\end{equation}
where the constants $\alpha,\beta,\gamma$ are given by $\alpha\equiv\frac{8E_nm}{\hbar^2}-\frac{24bm}{\delta\hbar^2}\beta\equiv\ -\frac{8am}{\hbar^2}+\frac{24bm}{\delta^2\hbar^2},\ \gamma\equiv-4L-4L^2-\frac{8bm}{\hbar^2\delta^3}$ and $B$ is found by \eqref{eq:8.3}. For the $n=1$ excited wave function with 2S energy level, B converges to .156816, and \eqref{eq:wavefunction 2} yields the $c\overline{c}$ meson wave function amplitude and probability density.

\begin{figure}[H]
 \vspace{0ex}
\centering
\includegraphics[width=.4\textwidth]{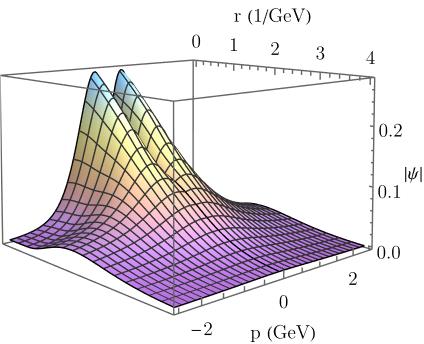}
\qquad
\includegraphics[width=.4\textwidth]{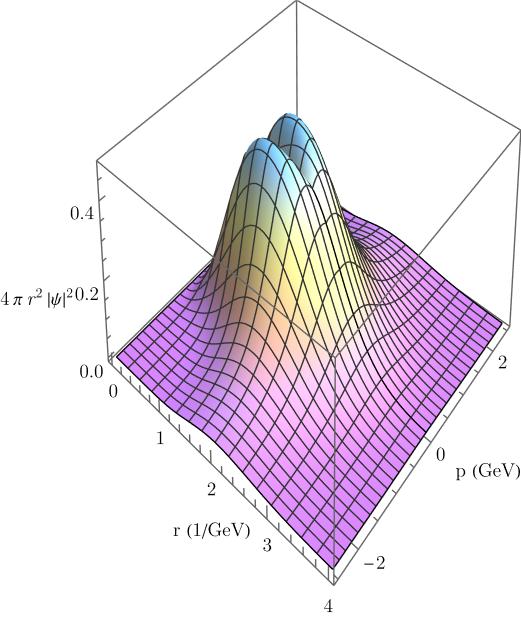}
\caption{2S $cc$ meson wave function probability amplitude(left) and density(right).\label{fig:6}}
\end{figure}

For the $n=1$ excited wave function with 2P energy level, $B$ converges to .0121014, and \eqref{eq:wavefunction 2} yields the $c\overline{c}$ meson wave function amplitude and probability density:

\begin{figure}[H]
\centering
\includegraphics[width=.4\textwidth]{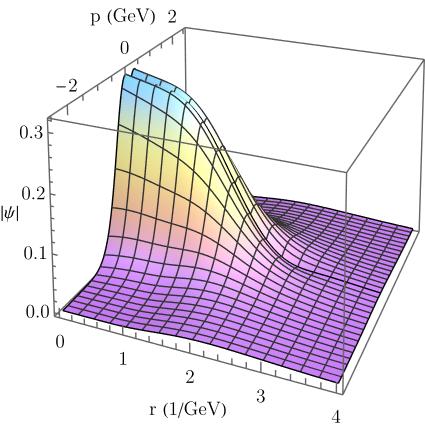}
\qquad
\includegraphics[width=.4\textwidth]{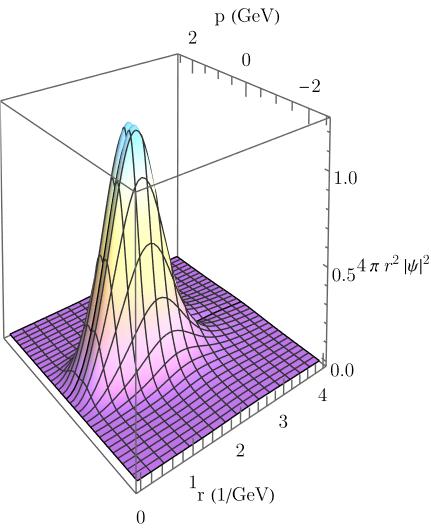}
\caption{2P $cc$ meson wave function probability amplitude(left) and density(right).\label{fig:7}}
\end{figure}

Using the methods described before, we also present the $n=2$ excited wave function amplitude and density for the 3S energy level, where $B$ converges to .0800241, and the $n=3$ excited wave function amplitude and density for the 4S energy level, where $B$ converges to .0003655.

\begin{figure}[H]
\centering
\includegraphics[width=.4\textwidth]{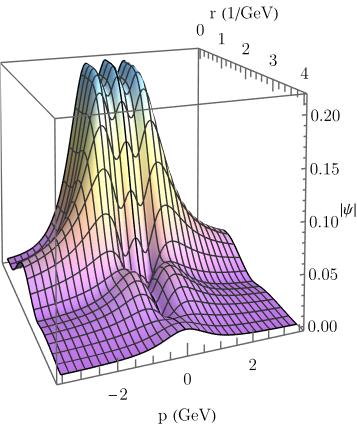}
\qquad
\includegraphics[width=.4\textwidth]{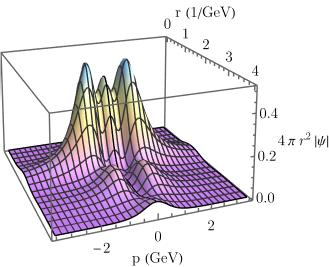}
\caption{3S $cc$ meson wave function probability amplitude(left) and density(right).\label{fig:8}}
\end{figure}

\begin{figure}[H]
\centering
\includegraphics[width=.4\textwidth]{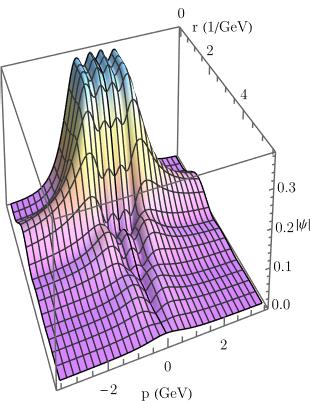}
\qquad
\includegraphics[width=.4\textwidth]{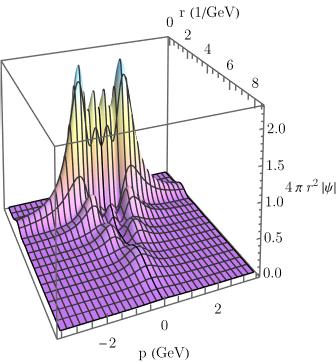}
\caption{4S $cc$ meson wave function probability amplitude(left) and density(right).\label{fig:9}}
\end{figure}

We also present graphs of each wave function density previously presented and how they change with varying radial momentum: \\

\begin{figure}[htbp]
\centering
\includegraphics[width=.4\textwidth]{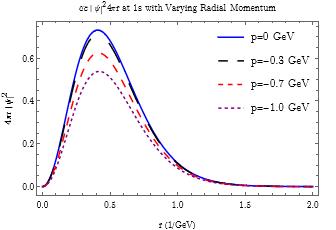}
\qquad
\includegraphics[width=.4\textwidth]{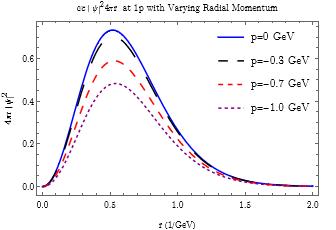}
\caption{$cc$ probability density, 1S (left) and 1P (right).\label{fig:10}}
\end{figure}
\begin{figure}[htbp]
\centering
\includegraphics[width=.4\textwidth]{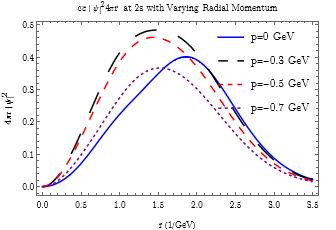}
\qquad
\includegraphics[width=.4\textwidth]{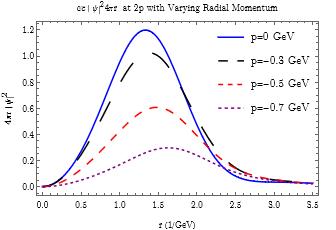}
\caption{$cc$ probability density, 2S (left) and 2P (right).\label{fig:11}}
\end{figure}
\begin{figure}[H]
\centering
\includegraphics[width=.4\textwidth]{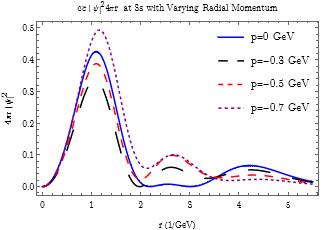}
\qquad
\includegraphics[width=.4\textwidth]{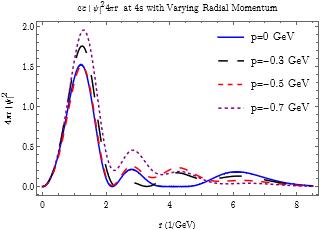}
\caption{$cc$ probability density, 3S (left) and 4S (right).\label{fig:12}}
\end{figure}

From these graphs, it is noted that a change in radial momentum actively changes the possible maximum relative quark anti-quark distance, suggesting that there is an upper limit of existence for charm-anticharm mesons that depends on the kinetic energy. \\

\bigskip

\section{Conclusion}
Through use of the alternative to the convolution and integral transforms, it can be seen that the Schrodinger equation in the quantum phase space representation can be half-transformed by its position operator to represent a collective variable. This new representation of the Schrodinger equation can be manipulated and approximated using power series to put into a hyper-geometric form. This procedure can be generalized for any potential of polynomial and reciprocal terms.\\

This form allows us to solve the phase space Schrodinger equation for its wave function and energy eigenvalues using the Nikiforov-Uvarov method. Using the traditional transformed phase space Schrodinger equation results in complex energy eigenvalues. These complex eigenvalues are suggested to arise from a rotational transform in the complex eigenvalues, which then to not add to result on the real axis. By analyzing the wave function expressions, a common exponential factor, $e^{-\frac{2i{p_r}r}{\hbar}}$, is found, corresponding to a phase space rotational transformation or phase shift. By creating an ansatz that the wave function is $\psi\left(r,{p_r}\right)=e^{-\frac{2i{p_r}r}{\hbar}}\ \Omega(r,{p_r})$, we can generate a Schrodinger equation describing $\Omega(r,{p_r})$ that eliminates all of the imaginary and momentum terms, resulting in real energy eigenvalues. \\

This particular method of transforming half of the domain, using a phase space rotation, then invoking a collective variable substitution such that the NU method could be used, may be a more general approach that was only under specific conditions here. We name this method the Half Transform Ansatz (HTA), and will continue with further investigation in upcoming papers.  \\

Fitting our energy eigenvalue equation with respect to experimental data describing the mass spectra of various mesons, we see that our model is more accurate than other theoretical models and shows a general trend that an increase in quantum numbers $n, L$ results in an increase in the mass spectrum. Analyzing our phase space wave functions for the indices $n=0,1,2,$ and $3$, we see that the radial momentum corresponds to an upper limit of existence in charm-anticharm mesons. 
\bigskip

\section{Future Developments and Acknowledgements}
\paragraph{Future Developments}
The main path of this work for the future is to adapt this framework for the Pauli and the Dirac QPSR equations, e.g, see section 3.4 in [5]. These developments will allow the HTA to be used to find the wave function and energy eigenvalues for relativistic particles with spin. We also continue our investigation into the HTA to explain the nature of the rotational factor that appeared along with the possibility that this example was only a specific result of a more general method. Additional developments also include adaptations of this work for baryons. These advancements and others will appear elsewhere.

\paragraph{Acknowledgements}
I would like to express my gratitude to the Gatton Academy, for their support in this research and funding to present this work at conferences. I also extend my appreciation to Valentino A. Simpao, at Western Kentucky University Physics and Astronomy, who helped nurture this project and gave me insight into the world of academia.

\bigskip

\begin{appendices}

\section{HOA and QPSR Hamiltonian Derivation}
\label{section:AppendixA}
This appendix gives additional information of the HOA and the construction of the Hamiltonian used in this paper. Beyond Refs. [3, 4, 5], previously developed HOA material by Simpao can be seen in Refs. [16, 17, 18]. It is also worth adding that the rigorous mathematical justifications for the HOA can be seen in [19]. These References in toto provide fundamental insight into the innerworks of the HOA and its applications. \\

Following Simpao [4, 5], below we present common Heaviside operational calculus methods and properties via Laplace transforms:

\begin{equation}
    \begin{aligned}
        L_{y\rightarrow\ z}\left[f\left(y\right)\right]=\int_{y_0}^{\infty}{f\left(y\right)e^{-yz}dy=\breve{f}\left(z\right)}
    \end{aligned}
\end{equation}
\begin{equation}
    \begin{aligned}
        L_{y\rightarrow\ z}^{-1}\left[\breve{f}\left(z\right)\right]=\frac{1}{2\pi i}\oint_{\partial}{\breve{f}\left(z\right)e^{yz}dz=f(y)}
    \end{aligned}
\end{equation}
\begin{equation}
\label{eq:A.3}
    \begin{aligned}
        {L}_{y\rightarrow\ z}^{-1}\left[\breve{f}\left(z\right)\right]=\frac{1}{2\pi i}\oint_{\partial}{\breve{f}\left(z\right)e^{yz}dz=f\left(y\right)=\breve{f}\left(D_y\right)U(y)}
    \end{aligned}
\end{equation}
\begin{equation}
\label{eq:A.4}
    \begin{aligned}
        L_{z\rightarrow\ y}^{-1}\left[{\breve{f}}_1\left(z\right){\breve{f}}_2\left(z\right)\right]=f_1\left(y\right)\ast f_2\left(y\right)=\int_{y_0}^{y}{f_1\left(y-u\right)f_2\left(u\right)du}
        \\
        ={\breve{f}}_1\left(D_y\right){\breve{f}}_2\left(D_y\right)U\left(y\right)={\breve{f}}_1\left(D_y\right)f_2\left(y\right)\ 
    \end{aligned}
\end{equation}
where U(y) is the Heaviside Unit Step function.
\begin{equation}
\label{eq:A.5}
    \begin{aligned}
         &{L}_{\left(y_1,\ldots,y_n\right)\rightarrow(z_1,\ldots,z_n)}[f(y_1,\ldots,y_n)] \\
        &=\int_{y_{0n}}^{\infty}{..n\ldots\int_{y_{0n}}^{\infty}{f\left(y_1,\ldots,y_n\right)e^{-\sum_{j=1}^{n}{y_jz_j}}}dy_1\ldots.dy_n}={\breve{f}}_1\left(z_1,\ldots,z_n\right)
    \end{aligned}
\end{equation}

\begin{equation}
\label{eq:A.6}
    \begin{aligned}
        L_{{(z}_1,\ldots,z_{n)}\rightarrow(y_1,\ldots,y_{n)}}^{-1}\left[\breve{f}\left(z_1,\ldots,z_n\right)\right]
        &=\left(\frac{1}{2\pi i}\right)^n\oint_{\partial^n}{\breve{f}\left(z_1,\ldots,z_n\right)e^{\sum_{j=1}^{n}{y_jz_j}}}dz_1\ldots..dz_n \\
        &=f(y_1,\ldots,y_n)
     \end{aligned}
\end{equation}
\begin{equation*}
    \begin{aligned}
        L_{{(z}_1,\ldots,z_{n)}\rightarrow(y_1,\ldots,y_{n)}}^{-1}\left[\breve{f_1}\left(z_1,\ldots,z_n\right)\breve{f_2}\left(z_1,\ldots,z_n\right)\right] \\
        =\ f_1\left(y_1,\ldots,y_n\right)\ \ \ \ \ \ \ast\ \ \ {\ \ \ f}_2\left(y_1,\ldots,y_n\right)\\
    \end{aligned}
    \begin{tikzpicture}[baseline=13.5pt]
        \hspace*{-100.5pt}
        \draw [decorate,decoration={brace,amplitude=5pt}](.6,0.0) -- (-.2,0.0);  
    \end{tikzpicture}
    \begin{tikzpicture}[baseline=27pt]
        \hspace*{-135pt}
        \node at (0,0) {$\left(p_1,\ldots,p_n\right)$};
    \end{tikzpicture}
\end{equation*}
\begin{equation}
\label{eq:A.7}
    \begin{aligned}
        &=\int_{y_{0_n}}^{x_n}{\ldots\ n\ldots}\int_{y_{0_1}}^{x_n}{f_1\left(y_1-y_1^\prime,\ \ldots,\ y_n-y_n^\prime\right)f_2\left(y_1^\prime,\ldots,y_n^\prime\right)} \\
       & ={\breve{f}}_1\left(\partial_{y_1},\ldots,\partial_{y_n}\right){\breve{f}}_2\left(\partial_{y_1},\ldots,\partial_{y_n}\right)U(y_1,\ldots,y_n) \\
       &={\breve{f}}_1\left(\partial_{y_1},\ldots,\partial_{y_n}\right)f_2\left(\partial_{y_1},\ldots,\partial_{y_n}\right)
    \end{aligned}
\end{equation} \\
\begin{equation*}
    \begin{aligned}
        f_1\left(x_1,\ldots,x_n;p_1,\ldots,p_n\right)\ \ \ \ \ \ \ \ \ \ast{\ \ \ \ \ \ \ \ f}_2(x_1,\ldots,x_n;p_1,\ldots,p_n),
    \end{aligned}
    \begin{tikzpicture}[baseline=1.5pt]
        \hspace*{-152.5pt}
        \draw [decorate,decoration={brace,amplitude=5pt}](.6,0.0) -- (-.2,0.0);  
    \end{tikzpicture}
    \begin{tikzpicture}[baseline=10pt]
        \hspace*{-192.5pt}
        \node at (0,0) {$\left(p_1,\ldots,p_n\right)$};
    \end{tikzpicture}
\end{equation*}
\begin{equation}
    \begin{aligned}
        =\int_{x_{0_n}}^{x_n}{\ldots n\ldots}\int_{0_1}^{x_1}\int_{p_{0_n}}^{p_n}{\ldots n\ldots}\int_{{p_0}_1}^{p_1}{f_1\left(x_1-x_1^\prime,\ldots,\ x_n-x_n^\prime;p_1-p_1^\prime,\ \ldots,\ p_n-p_n^\prime\right)}\\
        \times\ f_2(x_1^\prime,\ldots,\ x_n^\prime;p_1^\prime,\ldots,\ p_n^\prime)dx_1^\prime \ldots dx_n^\prime dp_1^\prime \ldots dp_n^\prime.
    \end{aligned}
\end{equation} \\
\begin{equation}
    \begin{aligned}
        L_{z\rightarrow\ y}\left[\breve{f}\left(az-b\right)\right]=\frac{1}{a}e^\frac{by}{a}f(\frac{y}{a})
    \end{aligned}
\end{equation}
\begin{equation}
    \begin{aligned}
        L_{(z_1,\ldots.,z_n)\rightarrow(y_1,\ldots.,y_n)}\left[\breve{f}\left(a_1z_1-b_1,\ldots,a_nz_n-b_n\right)\right]=\prod_{j=1}^{n}{\frac{1}{a}e^\frac{b_jy_j}{a_j}f(\frac{y_1}{a_1},\ldots,\frac{y_n}{a_n})}
    \end{aligned}
\end{equation}
From \eqref{eq:2.1}, the wave equation becomes	
\begin{equation}
    \begin{aligned}
        \hat{H}\binom{i\hbar\partial_{p_1}+\alpha_1x_1,\ldots,i\hbar\partial_{p_n}+\alpha_nx_n;}{-i\hbar\partial_{x_1}+\gamma_1p_1,\ldots,-i\hbar\partial_{x_n}+\gamma_np_n;t}&\psi\left(x_1,\ldots,x_n;p_1,\ldots,p_n;t\right)\\
        &=i\hbar\partial_t\psi(x_1,\ldots,x_n;p_1,\ldots,p_n;t).
    \end{aligned}
\end{equation}
Applying the convolution identity and multi-variable inverse transform of \eqref{eq:A.3}, the phase space convolution of \eqref{eq:A.4} and relation \eqref{eq:A.5} generates
\begin{equation}
    \begin{aligned}
        \left[L_{\binom{\left({\partial_{p_1},\ldots,\partial}_{p_n}\right)}{\rightarrow\left(p_1,\ldots.,p_n\right)}}^{-1}\left[L_{\binom{\left({\partial_{x_1},\ldots,\partial}_{x_n}\right)}{\rightarrow\left(x_1,\ldots.,x_n\right)}}^{-1}\left[\hat{H}\binom{i\hbar\partial_{p_1}+\alpha_1x_1,\ldots,i\hbar\partial_{p_n}+\alpha_nx_n;}{-i\hbar\partial_{x_1}+\gamma_1p_1,\ldots,-i\hbar\partial_{x_n}+\gamma_np_n;t}\right]\right]\right] \\
        \ast\ \ \ \ \ \ \ \ \ \ \ \psi\left(x_1,\ldots,x_n;p_1,\ldots,p_n;t\right)\\
        \\
        \equiv \hat{H}\binom{i\hbar\partial_{p_1}+\alpha_1x_1,\ldots,i\hbar\partial_{p_n}+\alpha_nx_n;}{-i\hbar\partial_{x_1}+\gamma_1p_1,\ldots,-i\hbar\partial_{x_n}+\gamma_np_n;t}\psi\left(x_1,\ldots,x_n;p_1,\ldots,p_n;t\right).                              
    \end{aligned}
    \begin{tikzpicture}[baseline=-4pt]
        \hspace*{-166.5pt}
        \draw [decorate,decoration={brace,amplitude=5pt}](.6,0.0) -- (-.2,0.0);  
    \end{tikzpicture}
    \begin{tikzpicture}[baseline=4pt]
        \hspace*{-206.5pt}
        \node at (0,0) {$\left(p_1,\ldots,p_n\right)$};
    \end{tikzpicture}
\end{equation}
Applying \eqref{eq:A.7}-\eqref{eq:A.6} with the convolution identities in \eqref{eq:A.3} yields
\begin{equation*}
    \begin{aligned}
        &L_{\binom{\left({p_1,\ldots,p}_n\right)}{\rightarrow\left({\bar{p}}_1,\ldots.,{\bar{p}}_n\right)}}\ \left[L_{\binom{\left({x_1,\ldots,x}_n\right)}{\rightarrow\left({\bar{x}}_1,\ldots.,{\bar{x}}_n\right)}}\ \left[\hat{H}\binom{i\hbar\partial_{p_1}+\alpha_1x_1,\ldots,i\hbar\partial_{p_n}+\alpha_nx_n;}{-i\hbar\partial_{x_1}+\gamma_1p_1,\ldots,-i\hbar\partial_{x_n}+\gamma_np_n;t}\psi\left(x_1,\ldots,x_n;p_1,\ldots,p_n;t\right)\right]\right] \\
        &\equiv L_{\binom{\left({p_1,\ldots,p}_n\right)}{\rightarrow\left({\bar{p}}_1,\ldots,{\bar{p}}_n\right)}}\\
        & \ \ \ \ \ \ \ \ \ \ \ \ \left[L_{\binom{\left({x_1,\ldots,x}_n\right)}{\rightarrow\left({\bar{x}}_1,\ldots,{\bar{x}}_n\right)}} \left[ L_{\binom{\left({\partial_{x_1},\ldots,\partial}_{x_n}\right)} {\rightarrow\left(x_1,\ldots.,x_n\right)}}^{-1}\left[L_{\binom{\left({\partial_{x_1},\ldots,\partial}_{x_n}\right)}{\rightarrow\left(x_1,\ldots.,x_n\right)}}^{-1}\left[\hat{H}\binom{i\hbar\partial_{p_1}+\alpha_1x_1,\ldots,i\hbar\partial_{p_n}+\alpha_nx_n;}{-i\hbar\partial_{x_1}+\gamma_1p_1,\ldots,-i\hbar\partial_{x_n}+\gamma_np_n;t}\right]\right]\right]\right] \\
        &\ \ \ \ \ \ \ \ \ \ \ \ \ \ \   \ast\ \ \ \ \ \ \ \ \ \ \ \psi\left(x_1,\ldots,x_n;p_1,\ldots,p_n;t\right)
    \end{aligned}
    \begin{tikzpicture}[baseline=52pt]
        \hspace*{-410.5pt}
        \draw [decorate,decoration={brace,amplitude=5pt}](.6,0.0) -- (-.2,0.0);  
    \end{tikzpicture}
    \begin{tikzpicture}[baseline=62pt]
        \hspace*{-452.5pt}
        \node at (0,0) {$\left(p_1,\ldots,p_n\right)$};
    \end{tikzpicture}
\end{equation*}
\begin{equation}
    \begin{aligned}
        \ &=i\hbar\partial_t\check{\psi}\left({\bar{x}}_1,\ldots,{\bar{x}}_n;{\bar{p}}_1,\ldots,{\bar{p}}_n;t\right) \\
        &\equiv\ \hat{H}\binom{i\hbar{\bar{p}}_1+\alpha_1x_1,\ldots,i\hbar{\bar{p}}_n+\alpha_nx_n;}{-i\hbar{\bar{x}}_1+\gamma_1p_1,\ldots,-i\hbar{\bar{x}}_n+\gamma_np_n;t}\check{\psi}\left({\bar{x}}_1,\ldots,{\bar{x}}_n;{\bar{p}}_1,\ldots,{\bar{p}}_n;t\right)&i\hbar\partial_t\check{\psi}\left({\bar{x}}_1,\ldots,{\bar{x}}_n;{\bar{p}}_1,\ldots,{\bar{p}}_n;t\right).
    \end{aligned}
\end{equation}
Hence, the wave function in phase space may be analytically expressed in exact quadrature. By inverse transforming the solution $\check{\psi}\left({\bar{x}}_1,\ldots,{\bar{x}}_n;{\bar{p}}_1,\ldots,{\bar{p}}_n;t\right)$, we have
\begin{equation}
    \begin{aligned}
        \psi_{r\ phasespace}\left(r;{p_r};t\right)={L^{-1}}_{\binom{\bar{r}\rightarrow r;}{\bar{{p_r}}\rightarrow {p_r}}}\begin{pmatrix}{e^{-\frac{i}{\hbar}\int_{0}^{t}H \begin{pmatrix}
        i\hbar{\bar{p}}_1+\alpha_1x_1,\ldots,i\hbar{\bar{p}}_n+\alpha_nx_n;\\ 
        -i\hbar{\bar{x}}_1+\gamma_1p_1,\ldots,-i\hbar{\bar{x}}_n+\gamma_np_n;t\prime)\end{pmatrix} dt\prime}}\\{\times\ {\breve{\psi}}_{0r\ phasespace}\left(\bar{r};{\bar{p}}_r;t=0\right)}\
        \end{pmatrix}.
    \end{aligned}
\end{equation} \\
We now begin to the Hamiltonian using the QPSR operators [3] and the definition 
\begin{equation}
    \begin{aligned}
        \hat{H}=\hat{T}+\hat{V},
    \end{aligned}
\end{equation}
where 
\begin{equation}
    \begin{aligned}
        \hat{T}=\sum^{N}{-\frac{\hbar^2}{2m}\nabla^2}\ =\ \sum^{N}{-\frac{\hbar^2}{2m}\left(\frac{\partial^2}{\partial x^2}+\frac{\partial^2}{\partial y^2}+\frac{\partial^2}{\partial z^2}\right)}.
    \end{aligned}
\end{equation}
$\hat{V}$ is dependent on the system’s potential. Working in the 1D space of $r$, this becomes   
\begin{equation}
    \begin{aligned}
        \hat{T}=\sum^{N}{-\frac{\hbar^2}{2m}\nabla^2}\ =\ \sum^{N}{-\frac{\hbar^2}{2m}\left(\frac{\partial^2}{\partial r^2}\right)}.
    \end{aligned}
\end{equation}
The momentum operator in the position basis is
\begin{equation}
    \begin{aligned}
        \hat{p}=-i\hbar\nabla\ =\left(-i\hbar\frac{\partial}{\partial r}\right).
    \end{aligned}
\end{equation}
Thus, for a one-dimensional system, 
\begin{equation}
    \begin{aligned}
        \hat{H}=\frac{{\hat{p}}^2}{2m}+V(r).
    \end{aligned}
\end{equation}
For a 3-dimensional system, we transform the coordinates of $\nabla$ via 
\begin{equation}
    \begin{aligned}
        \ x=r\sin\theta cos\phi,\ y&=r\sin\theta sin\phi,\ z=r\cos\theta \\
        \ r^2=x^2+y^2+z^2,\ \tan^2{\theta}&=\frac{x^2+y^2}{z^2},\ \tan\phi=\frac{y}{x}.
    \end{aligned}
\end{equation}
Using these equations, we find that 
\begin{equation}
    \begin{aligned}
        \frac{\partial r}{\partial x}=\sin\theta\cos\phi,\ \frac{\partial r}{\partial y}&=\sin\theta\sin\phi,\ \ \frac{\partial r}{\partial z}=\cos\theta \\
        \frac{\partial\phi}{\partial x}=-\frac{1sin\phi}{rsin\theta},\ \ \frac{\partial\phi}{\partial y}&=-\frac{cos\phi}{rsin\theta},\ \ \frac{\partial\phi}{\partial z}=0
    \end{aligned}.
\end{equation}
Using the Total Differential, we find that 
\begin{equation}
    \begin{aligned}
        \frac{\partial}{\partial x}\ =\ \ \frac{\partial r}{\partial x}\frac{\partial}{\partial r}\ +\ \frac{\partial\theta}{\partial x}\frac{\partial}{\partial\theta}\ +\ \frac{\partial\phi}{\partial x}\frac{\partial}{\partial\phi}\ &=\ sin\theta\ cos\phi\left(\frac{\partial}{\partial r}\right)+\frac{1}{r}cos\theta\ cos\phi\left(\frac{\partial}{\partial\theta}\right)-\frac{1}{r}\frac{sin\phi}{sin\theta}\left(\frac{\partial}{\partial\phi}\right), \\
        \frac{\partial}{\partial y}\ =\ \ \frac{\partial r}{\partial y}\frac{\partial}{\partial r}\ +\ \frac{\partial\theta}{\partial y}\frac{\partial}{\partial\theta}\ +\ \frac{\partial\phi}{\partial y}\frac{\partial}{\partial\phi}\ &=\ sin\theta\ sin\phi\left(\frac{\partial}{\partial r}\right)+\frac{1}{r}cos\theta\ sin\phi\left(\frac{\partial}{\partial\theta}\right)-\frac{1}{r}\frac{cos\phi}{sin\theta}\left(\frac{\partial}{\partial\phi}\right), \\
        \frac{\partial}{\partial z}\ =\ \ \frac{\partial r}{\partial z}\frac{\partial}{\partial r}\ +\ \frac{\partial\theta}{\partial z}\frac{\partial}{\partial\theta}\ +\ \frac{\partial\phi}{\partial z}\frac{\partial}{\partial\phi}\ &=\ cos\theta\left(\frac{\partial}{\partial r}\right)-\frac{1}{r}sin\theta\left(\frac{\partial}{\partial\theta}\right).
    \end{aligned}
\end{equation}
Using these three equations, we can express $\nabla^2$ as
\begin{equation}
    \begin{aligned}
        \left(\frac{\partial^2}{\partial x^2}+\frac{\partial^2}{\partial y^2}+\frac{\partial^2}{\partial z^2}\right)=\frac{\partial^2}{\partial r^2}+\frac{1}{r^2}\frac{\partial^2}{\partial\theta^2}+\frac{1}{r^2\sin^2{\theta}}\frac{\partial^2}{\partial\phi^2}+\frac{2}{r}\frac{\partial}{\partial r}+\frac{cot\theta}{r^2}\frac{\partial}{\partial\theta}.
    \end{aligned}
\end{equation}
Separating this into $r, \theta$, and $\phi$:
\begin{equation}
    \begin{aligned}
        \left(\frac{\partial^2}{\partial x^2}+\frac{\partial^2}{\partial y^2}+\frac{\partial^2}{\partial z^2}\right)=\frac{1}{r^2}\frac{\partial}{\partial r}\left(r^2\frac{\partial}{\partial r}\right)+\frac{1}{r^2}\hat{\Omega}\left(\theta,\ \phi\right),\ 
    \end{aligned}
\end{equation}
\begin{equation}
    \begin{aligned}
        \hat{\Omega}\left(\theta,\ \phi\right)\equiv\frac{1}{sin\theta}\left(\frac{\partial}{\ \partial\theta}\left(sin\theta\frac{\partial}{\partial\theta}\right)\right)+\frac{1}{\sin^2{\theta}}\left(\frac{\partial^2}{\partial\phi^2}\right).
    \end{aligned}
\end{equation}
Noticing that $\hat{\Omega}\left(\theta,\ \phi\right)$ is a function of angular dependency, we investigate orbital angular momentum. By definition,
\begin{equation}
    \begin{aligned}
        \vec{L}=\vec{r} \times\ \vec{p}=\left(yp_z-zp_y,\ zp_x-xp_z,\ xp_y-yp_x\right).
    \end{aligned}
\end{equation}
Thus,
\begin{equation}
    \begin{aligned}
        {\hat{L}}_z=-i\hbar\frac{\partial}{\partial\phi},\ {\hat{L}}_y=-i\hbar\left(\cos\phi\frac{\partial}{\partial\theta}-\cot\theta\ \sin\phi\frac{\partial}{\partial\phi}\right),\ {\hat{L}}_x=-i\hbar\left(-\sin\phi\frac{\partial}{\partial\theta}-\cot\theta\ \cos\phi\frac{\partial}{\partial\phi}\right)
    \end{aligned}
\end{equation}
Let us introduce a new operator ${\hat{L}}^2\equiv{\hat{L}}_x^2+{\hat{L}}_y^2+{\hat{L}}_z^2$. We can then see that 
\begin{equation}
    \begin{aligned}
        {\hat{L}}^2&=-\hbar^2\left(\frac{\partial^2}{\partial\theta^2}+\cot\theta\frac{\partial}{\partial\theta}+\left(\cot^2{\theta+1}\right)\frac{\partial^2}{\partial\phi^2}\right) \\
        &=-\hbar^2\left(\frac{1}{\sin{\theta}}\frac{\partial}{\partial\theta}\left(\sin{\theta}\frac{\partial}{\partial\theta}\right)+\frac{1}{\sin^2{\theta}}\frac{\partial^2}{\partial\phi^2}\right) \\
        &=-\hbar^2\hat{\Omega}(\theta,\ \phi).
    \end{aligned}
\end{equation}
Now we can say,
\begin{equation}
    \begin{aligned}
        \left(\frac{\partial^2}{\partial\ x^2}+\frac{\partial^2}{\partial\ y^2}+\frac{\partial^2}{\partial\ z^2}\right)=\frac{1}{r^2}\frac{\partial}{\partial r}\left(r^2\frac{\partial}{\partial r}\right)+\frac{1}{r^2}\hat{\Omega}\left(\theta,\ \phi\right)=\frac{1}{r^2}\frac{\partial}{\partial r}\left(r^2\frac{\partial}{\partial r}\right)-\frac{{\hat{L}}^2}{{r^2\hbar}^2},
    \end{aligned}
\end{equation}
so $\hat{T}$ becomes
\begin{equation}
    \begin{aligned}
        \hat{T}&=\ \sum^{N}{-\frac{\hbar^2}{2m}}\left(\frac{1}{r^2}\frac{\partial}{\partial r}\left(r^2\frac{\partial}{\partial r}\right)-\frac{{\hat{L}}^2}{r^2\hbar^2}\right) \\
        &=\sum^{N}{-\frac{\hbar^2}{2m}\frac{2}{r}\frac{\partial}{\partial r}-\frac{\hbar^2}{2m}\frac{\partial^2}{\partial r^2}+\frac{{\hat{L}}^2}{2mr^2}} \\
        &=\sum^{N}{\frac{-i\hbar{\hat{p}}_r}{m\hat{r}}+\frac{{\hat{p}}_r^2}{2m}+\frac{{\hat{L}}^2}{2m{\hat{r}}^2}}.
    \end{aligned}
\end{equation}
Since the eigenvalues of $L^2$ is $\hbar^2l(l+1)$, where $l$ is the orbital quantum number, we can rewrite this as 
\begin{equation}
    \begin{aligned}
        \hat{T}=\sum^{N}{\frac{-i\hbar{\hat{p}}_r}{m\hat{r}}+\frac{{\hat{p}}_r^2}{2m}+\frac{\hbar^2l(l+1)}{2m{\hat{r}}^2}}.
    \end{aligned}
\end{equation}
Now we have the 3D Hamiltonian in terms of $(\hat{r},\hat{p})$:
\begin{equation}
    \begin{aligned}
        \hat{H}=\sum^{N}{\frac{-i\hbar{\hat{p}}_r}{m\hat{r}}+\frac{{\hat{p}}_r^2}{2m}+\frac{\hbar^2l(l+1)}{2m{\hat{r}}^2}}+\hat{V}(r).
    \end{aligned}
\end{equation}

\bigskip 

\section{Time Dynamics of the 1D Confinement Wave Function}
This appendix presents the formulation of the time dynamics of the 1D quark confinement wave function via Fourier transform from the energy to time domain.\\

The energy of a particle written in the frequency domain is 
\begin{equation}
    \begin{aligned}
        E=hv
    \end{aligned}
\end{equation}
The non-normalized wave function is given from \eqref{eq:3.13} as
\begin{equation}
\label{eq:B.2}
    \begin{aligned}
        \Psi\left(r,p\right)=\ c_1Ai\left[\left(\frac{p^2}{2m}+br-E\right)\omega^{-\frac{1}{3}}\ \right].
    \end{aligned}
\end{equation}

Using the integral representation of the Airy $Ai$ function as stated in [9], \eqref{eq:B.2} can be written as
\begin{equation}
    \begin{aligned}
        \Psi\left(r,p\right)=\ c_1\int_{-\infty}^{\infty}e^{i\left(\frac{z^3}{3}\ +\ {\ z\left(\frac{1}{w}\right)}^\frac{1}{3}\left(\frac{p^2}{2m}+br-hv\right)\right)}dz.
    \end{aligned}
\end{equation}
Thus, the inverse Fourier transform of wave function from frequency to time domain in integral form is
\begin{equation}
    \begin{aligned}
        \frac{c_1}{2\pi}\int_{-\infty}^{\infty}\left(\int_{-\infty}^{\infty}e^{i\left(\frac{z^3}{3}\ +\ {\ z\left(\frac{1}{w}\right)}^\frac{1}{3}\left(\frac{p^2}{2m}+br-hv\right)\right)}dz\right)\ e^{itv}\ dv \\
        =\frac{c_1}{2\pi}\int_{-\infty}^{\infty}\left(\int_{-\infty}^{\infty}{e^{\frac{ip^2}{2m}\left(\frac{1}{w}\right)^\frac{1}{3}}e^{ibr\left(\frac{1}{w}\right)^\frac{1}{3}z}e^\frac{iz^3}{3}e^{ih\left(\frac{1}{w}\right)^\frac{1}{3}zv}}dz\right)\ e^{itv}\ dv.
    \end{aligned}
\end{equation}
Making the substitution $u=\ -h\left(\frac{1}{w}\right)^\frac{1}{3}z$,
\begin{equation}
    \begin{aligned}
        =\frac{c_1}{2\pi}\int_{-\infty}^{\infty}\left(\left(-h\left(\frac{1}{w}\right)^\frac{1}{3}\right)^{-1}\int_{-\infty}^{\infty}{e^{-\frac{ip^2u}{2hm}-\frac{ibru}{h}-\frac{iu^3w}{3h^3}}e^{-iuv}}du\right)\ e^{itv}\ dv.
    \end{aligned}
\end{equation}
Using the integral definition of the Fourier transform, this can be rewritten as 
\begin{equation*}
    \begin{aligned}
    c_1\ F_{(v\rightarrow\ t)}^{-1}\left[F_{\left(u\rightarrow\ v\right)}\left[\left(-h\left(\frac{1}{w}\right)^\frac{1}{3}\right)^{-1}e^{-\frac{ip^2u}{2hm}-\frac{ibru}{h}-\frac{iu^3w}{3h^3}}\right]\right] 
    \end{aligned}
\end{equation*}

\begin{equation}
    \begin{aligned}
        &=c_1\left(-\frac{1}{h}\left(\frac{1}{w}\right)^{-\frac{1}{3}}e^{-\frac{ip^2u}{2hm}-\frac{ibru}{h}-\frac{iu^3w}{3h^3}}\right)_{u\rightarrow\ t} \\
        &=c_1\left(-\frac{1}{h}\left(\frac{1}{w}\right)^{-\frac{1}{3}}e^{-\frac{ip^2t}{2hm}-\frac{ibrt}{h}-\frac{it^3w}{3h^3}}\right)\\
        &=c_1\left(-\frac{1}{2\pi\hbar}\left(\frac{m}{b^2\hbar^2}\right)^\frac{1}{9}e^{-\frac{ip^2t}{4\pi\hbar\ m}-\frac{ibrt}{2\pi\hbar}-\frac{it^3\left(\frac{m}{b^2\hbar^2}\right)^\frac{1}{3}}{24\pi^3\hbar^3}}\right).
    \end{aligned}
\end{equation}
Thus,
\begin{equation}
    \begin{aligned}
        \Psi\left(r,p,t\right)=c_1\left(-\frac{1}{2\pi\hbar}\left(\frac{m}{b^2\hbar^2}\right)^\frac{1}{9}e^{-\frac{ip^2t}{4\pi\hbar\ m}-\frac{ibrt}{2\pi\hbar}-\frac{it^3\left(\frac{m}{b^2\hbar^2}\right)^\frac{1}{3}}{24\pi^3\hbar^3}}\right)
    \end{aligned}
\end{equation}
\bigskip

\end{appendices}



\begin{thebibliography}{99}
\bibitem{a}
Augustin, J. -E. et al.,
\emph{Discovery of a Narrow Resonance in 
$e^+e^-$
 Annihilation},
\emph{Phys. Rev. Lett.} {\bf 33},  (1974) pg.1406-1408, 
\quad https://doi.org/10.1103/PhysRevLett.33.1406


\bibitem{b}
Eichten, E.; Gottfried, K.; Kinoshita, T.; Kogut, J. B.; Lane, K. D.; Yan, T. M.,
\emph{Spectrum of charmed quark-antiquark bound states},
\emph{Phys. Rev. Lett.} {\bf 34}, (1976) pg.369-372 \quad https://doi.org/10.1103/PhysRevLett.36.1276

\bibitem{c}
Torres-Vega, G.; Frederick, J.H.,
\emph{A quantum-mechanical representation in phase space.},
\emph{J. Chem. Phys.} {\bf 98} (1993) pg.3103-3120 \quad https://doi.org/10.1063/1.464085

\bibitem{d}
Simpao, V. A., 
\emph{Real wavefunction from Generalized Hamiltonian Schrodinger Equation in quantum phase space via HOA (Heaviside Operational Ansatz): Exact analytical results.},
\emph{J. of Math. Chem.} {\bf 52} (2014) pg.1136-1155 \quad https://doi.org/10.1007/s10910-014-0332-2

\bibitem{e}
Simpao, V. A., 
\emph{Toward chemical applications of Heaviside operational Ansatz: exact solution of radial
Schrodinger equation for nonrelativistic N-particle system with pairwise 1/rij radial potential in quantum
phase space.
},
\emph{J. Math. Chem.} {\bf 45} (2009) pg.129-140 \quad https://doi.org/10.1007/s10910-008-9372-9

\bibitem{f}
Vallee, O. and Manuel, S., 
\emph{Airy Functions and Applications to Physics}, {Imperial College Press} (2004) \quad https://doi.org/10.1142/p345

\bibitem{g}
Aspnes, D., 
\emph{Electric-Field Effects on Optical Absorption near Thresholds in Solids}
 \emph{Phys. Rev.} {\bf 147} (1966) pg.554-566 \quad https://doi.org/10.1103/PhysRev.147.554

\bibitem{h}
Patrignani C., et al., 
\emph{Review of particle physics 2016-2017.}
 \emph{Chinese Physics C.} {\bf 40} (2016) \quad http://doi.org/10.1088/1674-1137/40/10/100001

 \bibitem{i}
Omugbe, E.; Osafile, O. E.; Onyeaju, M. C., 
\emph{Mass spectrum of Mesons via the WKB approximation method.}
\emph{Advances in High Energy Physics} {\bf 2020} (2020) \quad https://doi.org/10.1155/2020/5901464

\bibitem{j}
Ibekwe, E.E.; Ngiangia, A.T.; Okorie, U.S. et al., 
\emph{Bound State Solution of Radial Schrodinger Equation for the Quark–Antiquark Interaction Potential.}
 \emph{Iran J Sci Technol Trans Sci} {\bf 44} (2020) pg.1191-1204 \quad https://doi.org/10.1007/s40995-020-00913-4

\bibitem{k}
Inyang, E. P.; Ntibi, J. E.; Ibekwe, E. E.; William, E. S., 
\emph{Approximate solutions of D-dimensional Klein–Gordon equation with Yukawa potential via Nikiforov–Uvarov method.}
 \emph{Indian Journal of Physics} {\bf 95} (2021) pg.2733-2739 \quad https://doi.org/10.1007/s12648-020-01933-x

\bibitem{l}
Rani, R.; Bhardwaj, S. B.; Chand, F., 
\emph{Mass Spectra of Heavy and Light Mesons Using Asymptotic Iteration Method.}
 \emph{Communications in Theoretical Physics.} {\bf 70} (2018) \quad https://doi.org/10.1088/0253-6102/70/2/179

\bibitem{m}
Abu-shady, M. and Khokha, E., 
\emph{Bound State Solutions of the Dirac Equation for the Generalized Cornell Potential Model.}
 \emph{International Journal of Modern Physics A.} {\bf 36} (2021) \quad https://doi.org/10.1142/S0217751X21501955

 \bibitem{n}
Abu-shady, M. and Fath-Allah, H. M., 
\emph{The Effect of Extended Cornell Potential on Heavy and Heavy-Light Meson Masses Using Series Method.}
 \emph{Journal for Foundations and Applications of Physics.} {\bf 6} (2019) \qquad https://arxiv.org/abs/1908.09131
 
 \bibitem{o}
Abu-Shady, M. and Ezz-Alarab, S. Y., 
\emph{Trigonometric Rosen–Morse Potential as a Quark–Antiquark Interaction Potential for Meson Properties in the Non-relativistic Quark Model Using EAIM.}
 \emph{Few-Body Systems.} {\bf 60} (2019) \qquad https://arxiv.org/abs/1905.05689
 
 \bibitem{p}
Simpao, V. A., 
\emph{HOA (Heaviside Operational Ansatz) revisited: recent remarks on novel exact solution methodologies in wavefunction analysis}
\emph{J. Math. Chem.} {\bf 50} (2012) pg.1931-1972 \quad https://doi.org/10.1007/s10910-012-0012-z

 \bibitem{q}
Simpao, V. A., 
\emph{In situ remarks on novel exact solutions of quantum dynamical systems: Heaviside operational ansatz in the quantum phase space representation at the generalised Hamiltonian-Lagrangian nexus}
\emph{Invited Book Chapter in ‘Focus on Quantum Mechanics” Nova Science Publishers, Inc.} (2011) 

 \bibitem{r}
Simpao, V. A., 
\emph{Recent Advances in Exact Analytical Wavefunction Methodologies’, [Invited monograph chapter in Theoretical Physics: Gravity, Magnetic Fields and Wave Functions Nova Publishing 2011]}

 \bibitem{s}
M. A. de Gosson and V. A. Simpao, 
\emph{Understanding the Schrödinger Equation Some [Non]Linear Perspectives, Chapter 8: From Classical to Quantum Physics: The Metatron}, {Nova Science Publishers, Inc} (2020), pg. 273.



\end{thebibliography}


\smallskip

\section*{Statements and Declarations}
The author declares that no funds, grants, or other support were received during the preparation of this manuscript. \\

Financial Interests: The author has no relevant financial or non-financial interests to disclose. \\

The independent author, Gabriel Nowaskie, did all work concerning this paper from the writing of the manuscript to the calculations and data analysis. The sole author, Gabriel Nowaskie, read and approved of this final manuscript.

\end{document}